\documentclass[conference,compsoc]{IEEEtran}
\IEEEoverridecommandlockouts

\usepackage{amsmath,amssymb,amsfonts}
\usepackage{amsmath}        %
\usepackage{algorithm}      %
\usepackage{algpseudocode}    %
\usepackage{amssymb}        %
\usepackage{graphicx}
\usepackage{textcomp}
\usepackage{multicol}
\usepackage{booktabs}
\usepackage{multirow}
\usepackage{tabularx}
\usepackage{makecell}
\usepackage{xcolor}
\usepackage{xspace}
\usepackage{todonotes}
\usepackage{threeparttable}
\usepackage{fancyhdr}
\usepackage{paralist}
\usepackage[normalem]{ulem}
\usepackage{bm} 
\newcommand{\ie}{i.e.,\xspace}
\newcommand{\eg}{e.g.,\xspace}

\usepackage{graphicx} %
\usepackage{subcaption}
\ifCLASSOPTIONcompsoc
  \usepackage[nocompress]{cite}
\else
  \usepackage{cite}
\fi

\ifCLASSINFOpdf
\else
\fi

\usepackage[utf8]{inputenc} %
\usepackage[T1]{fontenc}    %
\usepackage{xurl}
\usepackage[hidelinks]{hyperref}
\usepackage{booktabs}       %
\usepackage{amsfonts}       %
\usepackage{nicefrac}       %
\usepackage{microtype}      %
\usepackage{xcolor}         %
\usepackage{booktabs}
\usepackage{multirow}
\usepackage{comment}
\newcommand{\tool}{LambdaMark\xspace}
\newcommand{\wme}{$\mathcal{E}_{\theta}$\xspace}
\newcommand{\wmd}{$\mathcal{D}_{\phi}$\xspace}
\newcommand{\dash}{$\text{E}_{\text{s}}$\xspace}
\newcommand{\voc}{$\mathcal{V}_{\text{o}}$\xspace}

\newcommand{\madv}{$\mathcal{M}_{\text{adv}}$\xspace}
\newcommand{\dadv}{$\mathcal{D}_{\text{adv}}$\xspace}
\newcommand{\dorig}{$\mathcal{D}_{\text{orig}}$\xspace}
\newcommand{\dwm}{$\mathcal{D}_{\text{wm}}$\xspace}
\newcommand{\dwmsize}{$|\mathcal{D}_{\text{wm}}|$\xspace}
\newcommand{\newpar}[1]{\vspace{0.05cm}\noindent\textbf{#1.}\xspace}

\newcommand{\bitacc}{$\text{BRR}$\xspace}
\newcommand{\detacc}{$\text{Acc}_{\text{det}}$\xspace}

\newcommand{\audiomarkbench}{AudioSquareAttack\xspace}

\usepackage{marvosym}
\newcommand{\CorrAuth}{\textsuperscript{\normalfont\scriptsize\Letter}}

\begin{document}

\title{\tool: Semantic Audio Watermarking for Robustness and Radioactivity}

\author{
\IEEEauthorblockN{
Kexin Li\IEEEauthorrefmark{1}\CorrAuth,
Xiao Hu\IEEEauthorrefmark{1},
Ilya Grishchenko,
David Lie
}
\IEEEauthorblockA{
University of Toronto, Canada
}
\thanks{\IEEEauthorrefmark{1}Equal contributions}
\thanks{\CorrAuth Corresponding author}
}

\maketitle

\begin{abstract}
Recent advances in generative audio have made voice cloning increasingly effortless, enabling voice fraud, impersonation, and other forms of unauthorized use. A common attack finetunes a speech generation model on recordings of a target speaker, allowing the model to synthesize speech in that speaker's voice. Audio watermarking offers a promising defense by embedding detectable signals into audio. A practical watermark must satisfy two key properties: robustness, \ie it should remain detectable under audio manipulations and adversarial removal attacks, and radioactivity, \ie it should remain detectable after a downstream model is finetuned on watermarked audio. Existing audio watermarking methods typically embed signals into low-level representations, such as waveforms or spectrograms, which makes them vulnerable to signal-level manipulations and limits their transfer to downstream generative models.

We introduce \tool or \textcircled{$\lambda$}Mark---the first generic radioactive watermarking scheme.
Unlike all previous approaches, \tool achieves generic radioactivity by embedding multi-bit watermark information into semantic audio latent representations.
Consequently, our watermarks have semantic interpretation, and are thus more likely to be learned by a downstream model through finetuning.
\tool includes a lightweight watermark encoder to inject multi-bit message-dependent perturbations into semantic audio representations and a decoder to detect watermark presence and recover the embedded bit information.
Encoder and decoder are trained using a custom multi-component loss function that preserves fidelity of the watermarked audio, increases bit-level recovery rate, and improves robustness against common distortions and adversarial removal attempts.

Experiments show that \tool achieves near-perfect robustness under common distortions, with 99.92\% detection accuracy and 97.94\% bit recovery rate on LibriSpeech, and 99.88\% detection accuracy and 95.08\% bit recovery rate on VCTK as a cross-dataset transfer result. \tool is also the only watermark that is robust against all evaluated removal attacks. Furthermore, \tool exhibits general and robust radioactivity: its watermark remains detectable after finetuning diverse downstream audio generation models, including YourTTS, SemanticVocoder, and AudioLDM2, and remains robust to distortions and adversarial removal attacks even on the generated outputs of those finetuned models.

\end{abstract}

\IEEEpeerreviewmaketitle

\section{Introduction} \label{sec:intro}
Synthetic audio has become increasingly realistic and effortless to produce thanks to advances in generative audio models. Modern high-fidelity audio generation systems can perform tasks such as text-to-speech (TTS) synthesis, real-time voice conversion, and synthetic music generation~\cite{chuQwen2AudioTechnicalReport2024,kimiteamKimiAudioTechnicalReport2025,hu2026qwen3ttstechnicalreport}.
While these capabilities enable innovations and accessibility applications,  they also raise security and intellectual property risks. 
Synthetic speech can be used for impersonation, social engineering, and voice fraud~\cite{brewster2021voiceclone35m,guardian2024wppdeepfake,guardian2024hongkongdeepfake,park2024ai}. 
Furthermore, synthesized voice can be paired with facial deepfakes and video-synthesis tools to create convincing audio-visual impersonations, making fraudulent video calls or recordings more persuasive and catastrophic~\cite{guardian2024wppdeepfake,guardian2024hongkongdeepfake}.
At the same time, generative models trained on proprietary voices, songs, performances, or artistic recordings raise concerns for artists, voice actors, and dataset owners who want to verify whether their works have been used without authorization~\cite{RIAA_2024_Suno_Udio,iapp2023voiceactors,kanellopoulou2025unauthorized,guardian2025aigeneratedmusic}. 
AI-generated voices have already appeared in litigation over robocall deception and voter intimidation~\cite{AP_2024_BidenRobocallLawsuit,LWV_2025_KramerDefaultJudgment}. Further, voice actors have sued AI voice platforms for allegedly cloning and commercializing their voices without authorization~\cite{Reuters_2025_LovoVoiceActors}, and major record labels have sued AI music generators for allegedly training on copyrighted sound recordings without permission~\cite{Reuters_2024_SunoUdioLawsuits,Reuters_2025_UdioSettlement,Reuters_2025_WarnerSunoSettlement,Justia_2026_UdioMotionDismiss}. 
These examples motivate a common technical question: can we embed reliable signals for AI-generated content provenance and ownership verification into audio such that they remain detectable under realistic adversarial use? 

Audio watermarking is a defense for provenance and ownership verification. Existing methods embed inaudible hidden signals into audio so that a detector can later verify whether an audio sample contains the signal~\cite{audioseal,chenWavMarkWatermarkingAudio2024,singhSilentCipherDeepAudio2024,liuDetectingVoiceCloning2023,romanLatentWatermarkingAudio2025,liuGROOTGeneratingRobust2024,AudioMarkNet}. An audio watermark should satisfy two basic requirements. First, it should preserve \emph{fidelity}: the watermarked audio should sound natural and should not introduce audible artifacts. Second, it should be \emph{robust}: the watermark should survive common signal-level distortions and adversarial watermark removal attacks. However, if the goal is protection from voice cloning using generative audio models, a stronger property is required: \emph{radioactivity}, where a watermark embedded in the protected data can be learned by downstream models and can later be detected from their generated outputs. Importantly, radioactivity must also be robust: the watermark should remain detectable in downstream-generated samples even after post-finetuning manipulations or adversarial removal attempts. Further, watermarks with \emph{multi-bit} capacity allow traitor tracing (\ie~encoding an owner and identifying which protected source has been leaked). 

However, the pursuit of fidelity in recent watermarks has compromised their robustness, making radioactivity completely unachievable. Specifically, to avoid audible artifacts, modern audio watermarks often exploit psychoacoustic masking \cite{swansonRobustAudioWatermarking1998, spreadspectral}, embedding perturbations into time-frequency bins where stronger acoustic energy is present in the original audio content, making the watermark difficult for humans to perceive~\cite{swansonRobustAudioWatermarking1998,spreadspectral,audioseal,AudioMarkNet}. While this improves fidelity, it constrains the watermark to share consistent, localized residual patterns in the waveform or spectrogram. As a result, adaptive attackers can learn both the structure of the watermark and where it is likely to appear~\cite{li2025harmonicattackadaptivecrossdomainaudio}.

Recent watermark removal attacks expose this weakness. While signal-based and codec attacks test robustness against distortions~\cite{liuAudioMarkBenchBenchmarkingRobustness,oreillyDeepAudioWatermarks2025,ozerComprehensiveRealWorldAssessment2025}, a more recent adaptive attack, HarmonicAttack, demonstrates that a learning-based detector-free attacker can remove state-of-the-art audio watermarks without detector access, using only paired original and watermarked samples~\cite{li2025harmonicattackadaptivecrossdomainaudio}. Its transfer across watermarking schemes and audio domains suggests that existing watermarks share exploitable low-level acoustic structure induced by the psychoacoustic masking effect.
These findings motivate a new direction: rather than hiding watermarks as inaudible residual noise in the waveforms, audio watermarks should involve a semantic shift embedded into representations that are less directly exposed to residual-level removal attacks.

These semantic shifts have already been employed in a different domain---by the state-of-the-art image watermarking~\cite{wen2023treeringwatermarksfingerprintsdiffusion,ci2024ringidrethinkingtreeringwatermarking,li2025hmarkradioactivemultibitsemanticlatent}.
However, transitioning this result from the image domain to the audio has not been explored as far as we know. We believe this is due to a fundamental challenge---unlike image diffusion models that naturally have a semantic-latent space in the architecture~\cite{haas2024discoveringinterpretabledirectionssemantic}, audio generation models' semantic representations are less standardized and less directly accessible. This work, for the first time, explores how to use implicit semantic guidance for watermarking, specifically embedding watermarks in \emph{semantic audio latents} rather than directly perturbing the waveform or spectrogram residuals, as was done by all previous watermarking schemes.
Such semantic guidance is increasingly used in modern audio models. 
For instance, XiaoMi Dasheng scales masked audio encoder learning for general audio classification~\cite{xiaomidasheng}, 
and SemanticVocoder uses semantic latents to bridge audio understanding and waveform generation~\cite{xie2026semanticvocoderbridgingaudiogeneration}.

To introduce controlled semantic shifts in the latent space, \tool uses a novel semantic encoder--decoder architecture trained around a frozen audio encoder and semantic vocoder.
Embedding watermarks in semantic audio latents is challenging because latent perturbations must survive vocoder decoding and detector re-encoding while satisfying fidelity constraints to preserve perceptual audio quality and semantic consistency. \tool addresses these challenges with a uniform temporal broadcast encoder, a mean-pooling decoder, and a multi-objective loss function targeting fidelity, detection accuracy, and bit recovery simultaneously. 

As a result of such architecture and loss function design, the watermark exhibits high-level acoustic characteristics that are subsequently propagated into the decoded audio. Consequently, the embedded watermark is not tied to localized waveform residuals or fixed signal patterns, which are typically observed in existing audio watermarking schemes \cite{audioseal, chenWavMarkWatermarkingAudio2024, singhSilentCipherDeepAudio2024}, but rather encoded as a structured perturbation in the semantic latent manifold. During decoding, these latent-space modifications are transformed into globally distributed waveform variations. Such a design ensures \emph{robustness}: the watermark becomes harder to isolate or remove using conventional signal-processing operations or adaptive attacks that rely on identifying common residual shapes, localized embedding regions, or psychoacoustically masked perturbation patterns. 
\tool also supports \emph{radioactivity}: the watermark is embedded into the audio as a content-aligned semantic shift rather than a localized residual, thus, models finetuned on such protected audio treat the watermark as a structured, recurring feature in the training distribution rather than noise. 
As a result, the generated outputs inherit a recoverable watermark signal that a watermark detector can identify without access to the downstream model, and we empirically show that watermarks embedded by \tool survive finetuning.

\tool is, to the best of our knowledge, the first radioactive multi-bit semantic-guided audio watermarking scheme for adversarial robustness and ownership verification. Instead of embedding watermarks as low-level perturbations optimized mainly for inaudibility, \tool leverages semantic audio representations and injects multi-bit information into them. The resulting protected audio has high \emph{fidelity}, remaining perceptually close to the original, while the watermark is \emph{radioactive}---recoverable from adversarially-distorted or downstream-generated outputs. Because the watermark aligns with representations learned by downstream models, \tool enables black-box verification of whether protected audio was used for training. 
Moreover, \tool supports \emph{robust radioactivity}, meaning that the watermark remains detectable even when downstream-generated outputs undergo distortions or adversarial removal attempts, which is a more challenging attack scenario.
Furthermore, the \emph{multi-bit} capacity enables \tool to support more fine-grained attribution, such as traitor tracing.

In summary, this paper makes the following contributions:
\begin{itemize}
    \item We propose \tool, a semantic audio watermarking framework that embeds multi-bit information into semantic latent representations.
    \item We are the first to explore semantic-guided audio watermarking, and design a novel watermark encoder--decoder architecture and loss function for robustness, radioactivity, and fidelity.
    \item We show that, compared to baseline watermarking schemes, \tool is the only evaluated watermark that is robust against all the evaluated signal-level distortions, codec-based attacks, an optimization-based attack, and a learning-based adaptive attack.
    \item We demonstrate generalizable radioactivity through downstream finetuning on three diverse audio generation models, two datasets, and two finetuning methods, and show that the radioactive samples generated by downstream models are also robust against signal-level distortions and adversarial removal attacks.
\end{itemize}

\tool and all the code needed for reproduction will be released upon publication.

\section{Background} \label{sec:bg}

\begin{figure*}[th!]
    \centering
    \includegraphics[width=0.9\linewidth]{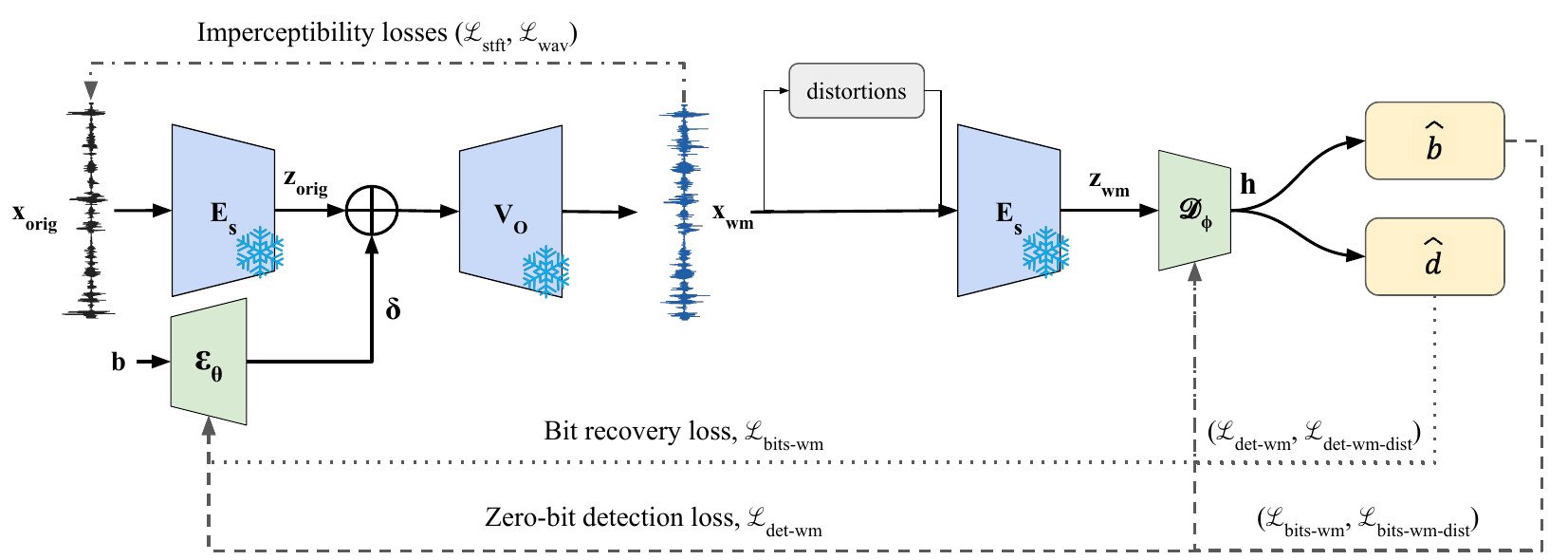}
    \caption{\tool's design overview. }
    \label{fig:overview}
\end{figure*}

\newpar{Audio Watermarking}
Audio watermarking aims to embed hidden information into audio signals in a way that is imperceptible to human listeners but can be detected by a decoder. AudioSeal~\cite{audioseal} employs a generator--detector architecture in which the generator directly synthesizes an additive watermark waveform in the time domain, while training objectives are designed to concentrate watermark energy within perceptually important time–frequency bins of the input signal, leveraging psychoacoustic masking to maintain perceptual fidelity. WavMark~\cite{chenWavMarkWatermarkingAudio2024} similarly constructs watermark signals through constrained perturbations applied to the waveform, explicitly optimizing for resilience against common distortions such as compression, filtering, and resampling while preserving imperceptibility under psychoacoustic constraints. AudioMarkNet~\cite{AudioMarkNet} further introduces an encoder–decoder framework that embeds message information into speech by injecting perturbations into selected spectral regions, particularly low-frequency and high-energy components, ensuring that the watermark is both perceptually masked and stable under downstream processing, including speaker adaptation and TTS generation.

Since these methods embed watermarks as waveform or spectrogram perturbations, their design is fundamentally tied to perceptual masking constraints \cite{swansonRobustAudioWatermarking1998, spreadspectral} that determine where and how watermark energy can be inserted to remain imperceptible to human listeners. This reliance on relatively structured and localized perturbation patterns makes them vulnerable to adaptive attacks \cite{li2025harmonicattackadaptivecrossdomainaudio} that learn to identify, model, and suppress the embedded residuals, particularly those induced by psychoacoustic masking-based strategies.

\newpar{Semantic Audio Representations and Generation}
Recent advances in generative audio modeling increasingly rely on semantic audio representations that encode high-level acoustic and structural information in compact latent spaces. Before large transformer-based audio generation models emerge, CLAP \cite{clap} learns a shared embedding space between audio and text through contrastive learning, enabling semantic-level audio understanding and text-audio alignment. AudioLDM2 \cite{audioldm2} extends this paradigm by combining semantic audio representations with latent diffusion, allowing high-quality text-conditioned audio generation through diffusion processes operating in compressed latent spaces.

Autoregressive semantic token generation models then emerge for audio synthesis. MusicGen \cite{musicGen} models discrete semantic audio tokens using transformer language models to generate coherent music conditioned on text or audio prompts. In parallel, neural vocoding models such as SemanticVocoder \cite{xie2026semanticvocoderbridgingaudiogeneration} focus on reconstructing waveforms from semantic acoustic representations, bridging semantic latent modeling and waveform synthesis. More recently, XiaoMi Dasheng \cite{xiaomidasheng} introduces a large audio encoder capable of jointly modeling speech, music, and environmental audio through unified semantic representations. Building upon this direction, MiDashengLM \cite{dinkel2026midashenglmefficientaudiounderstanding} extends semantic audio language modeling with stronger autoregressive generation capabilities and improved long-context audio modeling.

\begin{figure*}[th!]
    \centering
    \includegraphics[width=0.88\linewidth]{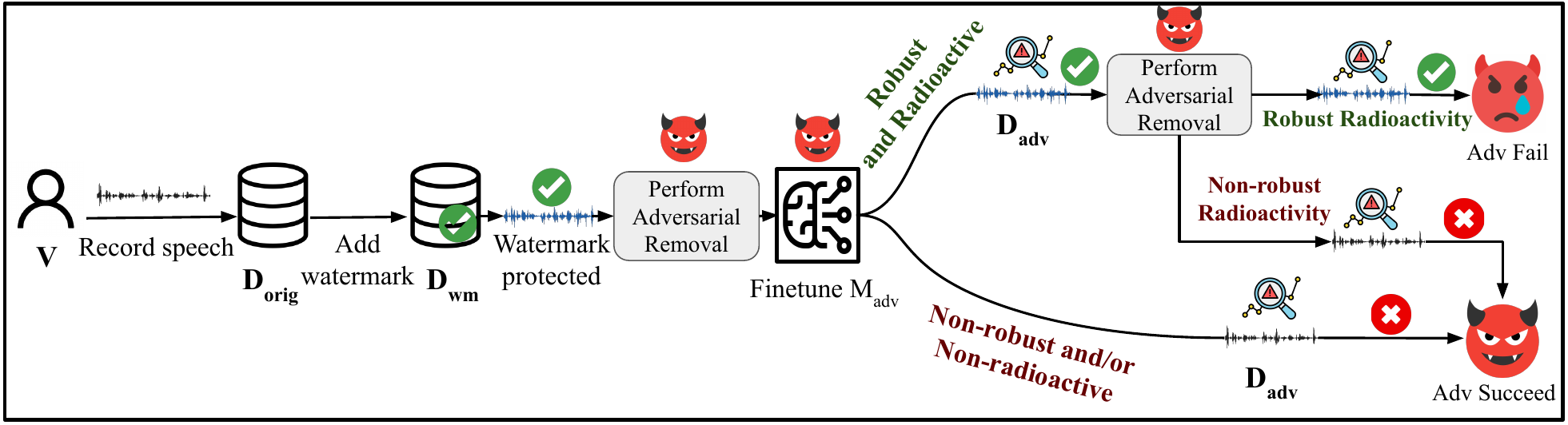}
    \caption{\tool protects \dorig from unauthorized voice cloning by embedding robust and radioactive watermarks. }
    \label{fig:threat}
\end{figure*}

\section{Design}
\label{sec:design}

\tool explores implicit semantic guidance for watermarking; specifically, it embeds watermarks into semantic audio latent representations via a novel encoder--decoder architecture and loss function.
As our watermark is embedded in the semantic latents of an audio waveform, the watermark identification is invariant to distortions as long as the audio's semantics are preserved, contributing to robustness against various distortions, adversarial attacks, and even downstream finetuning. 
See Figure~\ref{fig:overview} for a design overview.

\tool is a trainable watermark encoder--decoder that operates around a frozen semantic audio model backbone.
The semantic audio backbone needs to have a pre-trained semantic encoder \dash that can map audio into a semantically meaningful latent space, together with a decoder or vocoder that can synthesize audio from that latent representation.
During training, the watermark encoder \wme learns how to inject a recoverable signal into semantic latents before synthesis, while the watermark decoder \wmd learns how to identify and recover that signal after the audio has been synthesized, re-encoded, and potentially distorted.
This encoder--decoder design enables \tool to jointly optimize a custom loss for fidelity, detection, bit recovery, and robustness.

\subsection{Threat Model and Assumptions}

Voice cloning is a common and serious form of audio data misuse. In our threat model (See Figure~\ref{fig:threat}), we consider two parties: a victim $V$ (\eg a voice actor) and an adversary $Adv$ who seeks to clone $V$'s voice for malicious purposes, such as impersonation, blackmail, or intellectual property infringement. The victim protects their audio data \dorig by applying watermarking, producing the protected dataset \dwm, which may later become accessible to $Adv$. 

The adversary may obtain $V$'s voice samples through public downloads, leaked recordings, or fraudulent calls that are recorded without authorization. Through such operations, $Adv$ can obtain and leverage the protected audio \dwm to finetune a generative audio model \madv without permission and use the resulting model to synthesize audio for malicious activities. We assume an adversary who can manipulate and attempt adversarial removals on \dwm before finetuning or on \dadv after finetuning. The victim later obtains generated samples \dadv from \madv and can provide evidence to relevant authorities that the samples are derived from \dwm by demonstrating the existence of the watermark on samples in \dadv.
This setting is realistic because such generated samples are typically what first raises suspicion of misuse. We further assume that $V$ does not require access to \madv or its finetuning method. 

For the watermark to remain detectable under this challenging threat model, it must satisfy both robustness and robust radioactivity. Specifically, it must survive common distortions, downstream finetuning, and adversarial removal attempts regardless of whether the adversary applies them to \dwm \textit{before} finetuning or to \dadv \textit{after} finetuning.

\subsection{Watermark Encoder}
\label{sec:enc}
Unlike prior audio watermarking methods that embed watermark perturbations directly into the waveform \cite{audioseal} or into time-frequency representations \cite{chenWavMarkWatermarkingAudio2024}, our \emph{watermark encoder}, \wme, operates in the semantic latent space of a pre-trained audio encoder.  Given an N-bit message, \wme produces an additive residual $\delta$ that is injected into the semantic latents $z_{\text{orig}}$ produced by the chosen pre-trained semantic backbone (\S~\ref{sec:design:backbone}). 
\wme's input is only the message. This design aligns with our goal: to embed a watermark in audio's semantic latents, we induce a consistent semantic shift in the audio encoder \dash's latent space across all watermarked audio, making it easier for downstream models to learn, instead of adding potentially different perturbations to each waveform $x_\text{orig}$.
The watermarked audio is then synthesized by decoding the perturbed latent $z + \delta$ back into the waveform with the semantic vocoder,~\voc. By embedding in the latent space, the watermark depends on high-level audio content rather than waveform details, making it more robust to attacks. 

\wme is a lightweight three-layer MLP that maps the binary message to a single latent-dimensional direction. The N-bit message is first remapped from $\{0,1\}$ to $\{-1,+1\}$ to give the network a zero-centered input and then projected by the MLP into a vector $\tilde{f} \in \mathbb{R}^D$, where $D$ is the backbone latent dimension. We $\ell_2$-normalize the MLP output to obtain a unit direction $f = \tilde{f}/||\tilde{f}||_2$. 
This direction is then uniformly broadcast across all $T$ latent frames to form the residual $\delta \in \mathbb{R}^{D \times T}$. This uniform temporal broadcast is a deliberate design choice: it matches the mean-pooling operation in the watermark decoder so that the decoder reads the same ``average direction'' that \wme injected, concentrating the entire bit-budget into a single time-averaged direction.

A naive additive residual at a fixed magnitude would either be too weak on loud, spectrally rich audio (and thus undetectable) or too strong on quiet, sparse audio. To make the watermark amplitude adaptive, \wme scales the residual $\delta = \alpha \cdot \lVert z_{\text{orig}} \rVert \cdot f$ relative to the $\ell_2$ norm of the original latent, where $\lVert z_{\text{orig}} \rVert$ denotes the mean per-frame $\ell_2$ norm of the original latents and $\alpha$ is a bounded learnable scalar. This ensures that the perturbation magnitude scales with the energy of the semantic representation so that high-energy audio receives a stronger watermark signal. This design is analogous to psychoacoustically masked embedding in signal-domain watermarking~\cite{audioseal,AudioMarkNet}, but operates entirely in the semantic space, where the masking signal is given by the latent activation norm itself. 
We parameterize $\alpha = \alpha_{\max}\cdot\sigma(\theta_\alpha)$, where $\sigma$ is the sigmoid and $\theta_\alpha \in \mathbb{R}$ is a free parameter, so $\alpha \in (0, \alpha_{\max})$. The hard cap $\alpha_{\max}$ (the largest watermark-to-content magnitude ratio we allow) prevents the model from trivially maximizing bit recovery by inflating $\alpha$ at the expense of fidelity. 
The scale $\alpha$ is learned jointly with the watermark encoder and decoder, so the model can autonomously discover the right trade-off between detectability and fidelity within this bound.

\subsection{Watermark Decoder}
\label{sec:dec}
The \emph{watermark decoder}, \wmd, recovers the embedded message directly from the watermarked semantic latents $z_{\text{wm}}$, and it never sees the waveform. Operating in the same latent space where the watermark is embedded provides two advantages. First, it facilitates robustness to signal-level editing that preserves semantic representations, such as resampling, filtering, and noise signals. Second, a shared representation space between the watermark encoder \wme and decoder \wmd enables joint optimization of watermark embedding and recovery. 

The decoder mean-pools across the temporal dimension,
$\bar{z} = \frac{1}{T} \sum_{t=1}^{T} z \in \mathbb{R}^{B \times D}$,
mirroring \wme's uniform temporal broadcast strategy, where $B$ is the batch size and $D$ is the latent dimension. The pooled representation is then processed by an MLP consisting of two linear layers with GELU activations, producing a shared feature representation $h$. Two parallel prediction heads are attached to $h$. The first is a multi-bit decoding head that predicts the embedded message bits,
$\hat{b} = W_{\text{bit}} h + b_{\text{bit}} \in \mathbb{R}^{N}$,
where each bit is decoded by thresholding its corresponding logit at zero. The second is a zero-bit detection head,
$\hat{d} = W_{\text{det}} h + b_{\text{det}} \in \mathbb{R}$.
Both heads share the same feature representation $h$, allowing watermark presence detection and message recovery to benefit from a common latent representation.

\subsection{Loss Function} \label{sec:loss}
Our training objective jointly optimizes \wme (\S~\ref{sec:enc}) and \wmd (\S~\ref{sec:dec}) using the losses $\mathcal{L}$ that balance four goals: (1) fidelity of the audio with the embedded watermark ($\mathcal{L}_{\text{stft}}$, $\mathcal{L}_{\text{wav}}$), (2) zero-bit detection on watermarked audio ($\mathcal{L}_{\text{det-wm}}$), (3) bit-level message recovery on watermarked audio ($\mathcal{L}_{\text{bit-wm}}$), and (4) robustness to common signal-level distortions ($\mathcal{L}_{\text{det-wm-dist}}$, $\mathcal{L}_{\text{bit-wm-dist}}$). The total loss $\mathcal{L}_{\text{total}}$ is a weighted sum of the aforementioned losses:
\begin{equation*}
\begin{aligned}
\mathcal{L}_{\text{total}} =\;
  & \lambda_{\text{stft}} \mathcal{L}_{\text{stft}}
  + \lambda_{\text{wav}}  \mathcal{L}_{\text{wav}}
  + \lambda_{\text{bc}}   \mathcal{L}_{\text{bits-wm}} \\
  + & \lambda_{\text{bd}}   \mathcal{L}_{\text{bits-wm-dist}}
  + \lambda_{\text{dc}}   \mathcal{L}_{\text{det-wm}}
  + \lambda_{\text{dd}}   \mathcal{L}_{\text{det-wm-dist}}.
\end{aligned}
\end{equation*}

\textbf{Fidelity Losses.} To ensure fidelity, we use two losses. The first is a multi-resolution STFT loss enforcing the spectrum of $x_{\text{wm}}$ to remain close to that of $x_{\text{orig}}$:
\begin{equation*}
\begin{aligned}
\mathcal{L}_{\text{stft}}
  = \frac{1}{K}\!\sum_{k=1}^{K}
    \Bigg\lVert
      \log\!\big(1{+}|\text{STFT}_k(x_{\text{wm}})|\big)
      \\
      - \log\!\big(1{+}|\text{STFT}_k(x_{\text{orig}})|\big)
    \Bigg\rVert_{1}
\end{aligned}
\end{equation*}
The loss is computed over $K{=}4$ scales spanning short to long windows to capture differences at different scales. The 
$log$ compression prevents loud frames from dominating the loss. Since $\mathcal{L}_{\text{stft}}$ is phase-insensitive, we complement it with a waveform $\ell_1$ term to penalize gross phase or DC drift:
\begin{equation*}
\mathcal{L}_{\text{wav}}
  = \big\lVert x_{\text{wm}} - x_{\text{orig}} \big\rVert_{1}
\end{equation*}

\textbf{Bit Recovery Losses.}
At deployment, the detector only receives a \emph{waveform}, while the internal latents $z + \delta$ that \wme produces never leave the watermark embedder, since they are immediately consumed by \voc to synthesize $x_{\text{wm}}$. To recover the message, the detector must therefore re-encode the candidate waveform through \dash and run the decoding on the resulting latents. This re-encoding is not lossless:
\dash~$($\voc$(z + \delta)) \neq z + \delta$, because \voc is a lossy synthesis step. Training \wmd directly on $z + \delta$ mismatches with deployment. 

To match this condition during training, we synthesize the watermarked waveform $x_{\text{wm}}$ from the perturbed latents
$z + \delta$ via \voc and then re-encode it through \dash to obtain $z_{\text{wm}} = \text{\dash}(x_{\text{wm}})$, on which \wmd
reads bit logits $\hat{b}_{\text{wm}}$. 
Optimizing against $z_{\text{wm}}$ rather than $z + \delta$ directly forces \wme to embed a residual that survives the complete audio to latent synthesis and re-encoding round-trip, which is the exact path a deployed detector follows. 
To satisfy our robustness goal, a distorted variant additionally applies a random sampled signal-level distortion to $x_{\text{wm}}$ before re-encoding, yielding logits $\hat{b}_{\text{wm-dist}}$. Both terms are binary cross-entropy (BCE) against the ground-truth bits $b$:
\begin{equation*}
\mathcal{L}_{\text{bits-wm}} = \text{BCE}(\hat{b}_{\text{wm}}, b),
\qquad
\mathcal{L}_{\text{bits-wm-dist}}  = \text{BCE}(\hat{b}_{\text{wm-dist}},  b).
\end{equation*}
Only the original watermark term backpropagates through the full \wmd~$\rightarrow$~\dash~$\rightarrow$~\voc~$\rightarrow$~\wme path.
We detach the distorted term, $x_{\text{wm}}$, before applying distortion so that $\mathcal{L}_\text{bits-wm-dist}$ only backpropagates to \wmd. This design choice is based on cost, stability, and role separation considerations. First, not propagating the distorted term to \wme avoids a second backward pass through the vocoder ODE per step, which roughly doubles training time and peak memory. It also keeps non-differentiable distortions from injecting biased gradients into \wme, causing training instability. In our threat model, distortions occur after embedding, so treating robustness as a watermark decoder responsibility matches deployment and keeps the watermark encoder focusing on two existing objectives---fidelity and bit recovery.

\textbf{Zero-bit Detection Losses.}
The zero-bit head is trained as a binary classifier with watermarked latents as positives and original latents as negatives, with distorted watermarked inputs. Same as bit recovery, the positive latents are obtained by re-encoding $x_{\text{wm}}$ (and its distorted counterpart) through \dash, and the original watermark variant backpropagates through \wme, while the distorted variant trains only the detection head:
\begin{equation*}
\mathcal{L}_{\text{det-}\star}
  = \tfrac{1}{2}\!\left[
      \text{BCE}(\hat{d}_{\text{pos-}\star}, 1)
    + \text{BCE}(\hat{d}_{\text{neg-}\star}, 0)
    \right],
\end{equation*}
\begin{equation*}
\quad \star \in \{\text{wm},\,\text{wm-dist}\}.
\end{equation*}

\subsection{Pre-trained Backbone}
\label{sec:design:backbone}

To train our watermark encoder--decoder, \tool requires a suitable semantic backbone, including a pre-trained semantic encoder, which we denote as \dash that maps audio to frame-level semantic latents and a compatible decoder/vocoder, denoted as \voc that reconstructs audio waveforms from those semantic latents. %

Both components are kept frozen, but gradients still flow through them during the backpropagation, providing a learning signal.

The semantic backbone provides two advantages that are central to our design. First, \textit{semantic locality}.
Semantic audio representations are trained in \dash to capture high-level attributes such as phonetic content, prosody, and timbre. 

Watermark perturbations embedded in this latent space become associated with the semantic content of the audio rather than with specific waveform details. This makes the watermark more robust to re-synthesis and generation-based attacks, which are the primary threat considered in our setting. 
The second advantage is \textit{end-to-end differentiability}.  The perturbed latent representation $z + \delta$ is passed through \voc to synthesize a watermarked waveform $x_{\text{wm}}$, which is then re-encoded by \dash to obtain $z_{\text{wm}}$ before watermark decoding. This forms a completely differentiable pipeline. The watermark recovery losses (multi-bit and zero-bit detection BCE) are evaluated on $z_{\text{wm}}$, so their gradients propagate through the full round-trip, allowing \wme to directly learn perturbations that remain recoverable after the complete synthesis and re-encoding process. The fidelity losses (multi-resolution STFT and waveform $\ell_1$), in contrast, are evaluated directly on $x_{\text{wm}}$ and therefore back-propagate through \voc to \wme, but not through \dash's re-encoding path. This division of labor is intentional: bit recovery must survive the full round-trip a real detector would follow at inference time, while perceptual quality is a property of the synthesized waveform itself and does not require re-encoding to be measured.

\section{Evaluation} \label{sec:eval}

All experiments were performed on a machine with two Intel Xeon 6548Y CPUs with 512 GB RAM, and four Nvidia H100 GPUs with 96GB memory.
Unless otherwise specified, we set the watermark bit size to 32 and LoRA rank to 8.

\newpar{Dataset and Models}
In the evaluation, we instantiate Dasheng~\cite{xiaomidasheng} as the semantic encoder \dash, which maps 16 kHz raw audio to 768-dim frame-level semantic embeddings at 25 fps, and SemanticVocoder~\cite{xie2026semanticvocoderbridgingaudiogeneration} as the vocoder \voc, which synthesizes 24 kHz waveforms from those embeddings via a small number of ODE integration steps (See \S~\ref{sec:design:backbone}).

We conduct our evaluation on LibriSpeech~\cite{librispeech} and VCTK~\cite{yamagishi2019vctk}. LibriSpeech is a large-scale read-speech corpus, while VCTK is a multi-speaker English speech corpus containing recordings from 110 speakers with diverse accents. Both corpora are widely used in audio processing.
We include VCTK because our main competitor, AudioMarkNet~\cite{AudioMarkNet}, reports that one of its key limitations is no generalization to unseen speakers other than the 11 VCTK speakers on which they train their watermark model. Therefore, to compare with AudioMarkNet, we also evaluate \tool on VCTK. 
Despite that, we train \tool's watermark encoder--decoder only on LibriSpeech. 
All \tool evaluations on VCTK are therefore cross-dataset transfer evaluations.

We use the \texttt{train-clean-100} subset of LibriSpeech for watermark encoder--decoder training, and reserve 120 samples from the \texttt{test-clean} subset for validation.  
For each dataset, we construct an original audio set \dorig containing 2{,}500 samples that require protection. We then apply \tool to \dorig to obtain the watermarked version \dwm. These samples are not seen during watermark encoder--decoder training. For LibriSpeech, \dorig is drawn from non-overlapping samples in the \texttt{test-clean} subset. For VCTK, this separation is guaranteed by construction because VCTK is never used to train the watermark encoder--decoder.

We evaluate the radioactivity of \tool on three representative downstream audio generation models. First, we evaluate on YourTTS~\cite{yourtts}, a zero-shot multi-speaker text-to-speech and voice-conversion model built on VITS. This setting enables direct comparison with AudioMarkNet~\cite{AudioMarkNet}, whose radioactivity evaluation is limited to YourTTS.
We then extend the evaluation to two additional downstream architectures: SemanticVocoder~\cite{xie2026semanticvocoderbridgingaudiogeneration}, a semantic-guided audio generation model, and AudioLDM2~\cite{audioldm2}, a diffusion-based text-to-audio generation framework. By moving beyond the single YourTTS setting considered by AudioMarkNet, we test whether \tool's radioactive watermark remains effective across diverse audio generation pipelines rather than overfitting to one downstream model family and one dataset.
We assume \dwm does not have transcripts by default, so the adversary needs to transcribe \dwm before finetuning \madv. We use \texttt{openai/whisper-small} model~\cite{whisper} for transcription. 
After finetuning each \madv, we generate 100 samples (\dadv) from \madv and evaluate radioactivity and robustness on these generated samples. This allows us to determine whether the watermark signal remains detectable and robust after downstream model finetuning.

\newpar{Evaluation Stages} We evaluate \tool in \emph{two} stages. First, immediately after watermark embedding, we evaluate watermark effectiveness, fidelity, and robustness under common distortions and adversarial removal attacks. To reduce computational cost, we randomly select 100 samples from \dorig and their watermarked versions in \dwm for evaluation. Second, we evaluate radioactivity after adversarial finetuning. Here, \dwm serves as the downstream training set used by the adversary to finetune \madv. Unless otherwise specified, \madv is finetuned on the full \dwm of 2{,}500 samples. We also evaluate the robustness on \madv's generated samples \dadv in this stage. The size of \dadv is also set to 100.

\begin{table*}[th!]
\centering
\renewcommand{\arraystretch}{1.15}
\setlength{\tabcolsep}{8pt}
\caption{Comparison of watermark effectiveness, robustness, and fidelity with baseline methods. \tool is trained on LibriSpeech; therefore, all VCTK results of \tool are cross-dataset transfer results. }
\label{tab:distortion_results_comparison}
\begin{tabular}{lllcccccc|cc}
\toprule
\textbf{Distortions} & \textbf{Dataset} & \textbf{Watermark}
& \textbf{\detacc} & \textbf{\bitacc}
& \textbf{P~(\%)} & \textbf{R~(\%)} & \textbf{F1~(\%)}
& \textbf{AUC} & \textbf{ViSQOL} & \textbf{NISQA} \\
\midrule

\multirow{6}{*}{\textbf{Without}}
& \multirow{3}{*}{LibriSpeech}
& AudioSeal & 100.00 & 51.94 & 100.00 & 100.00 & 100.00 & 1.000 & \textbf{4.94} & 4.49 \\
& & WavMark & 100.00 & \textbf{100.00} & 100.00 & 100.00 & 100.00 & 1.000 & 4.66 & 4.52 \\
& & \tool & 100.00 & 99.75 & 100.00 & 100.00 & 100.00 & 1.000 & 4.44 & \textbf{4.66} \\
\cmidrule{2-11}
& \multirow{2}{*}{VCTK}
& AudioMarkNet & 93.50 & 83.54 & \textbf{100.00} & 87.00 & 93.05 & 0.935 & \textbf{4.60} & 4.54 \\
& & \tool & \textbf{100.00} & \textbf{98.13} & \textbf{100.00} & \textbf{100.00} & \textbf{100.00} & \textbf{1.000} & 4.26 & \textbf{4.81} \\

\midrule

\multirow{5}{*}{\textbf{With}}
& \multirow{3}{*}{LibriSpeech}
& AudioSeal & 91.13 & 50.22 & \textbf{100.00} & 82.25 & 83.69 & 0.911 & \textbf{4.28} & 3.44 \\
& & WavMark & 90.92 & 81.32 & \textbf{100.00} & 81.83 & 84.27 & 0.909 & 4.13 & 3.50 \\
& & \tool & \textbf{99.92} & \textbf{97.94} & 99.92 & \textbf{99.92} & \textbf{99.92} & \textbf{0.999} & 3.98 & \textbf{3.55} \\
\cmidrule{2-11}
& \multirow{2}{*}{VCTK}
& AudioMarkNet & 85.54 & 78.92 & 97.62 & 71.25 & 79.91 & 0.855 & \textbf{3.92} & 3.33 \\
& & \tool & \textbf{99.88} & \textbf{95.08} & \textbf{99.75 }& \textbf{100.00} & \textbf{99.88} & \textbf{0.999} & 3.74 & \textbf{3.65} \\

\bottomrule
\end{tabular}
\end{table*}

\newpar{Evaluation Metrics} To evaluate \tool{} and the baseline watermarks, we use metrics that measure watermark detection accuracy and audio quality preservation. 
For watermark detection, we report the detection accuracy (\detacc), bit recovery rate (\bitacc), precision (P), recall (R), F1 score, and area under the ROC curve (AUC). Detection accuracy measures the overall proportion of correctly classified samples, defined as \detacc~\( = \frac{TP + TN}{TP + TN + FP + FN}\). We compute P and R as \(\mathrm{P} = \frac{TP}{TP + FP}\) and \(\mathrm{R} = \frac{TP}{TP + FN}\), respectively.
We also report the F1 score, which balances P and R and is defined as \(\mathrm{F1} = \frac{2 \cdot \mathrm{P} \cdot \mathrm{R}}{\mathrm{P} + \mathrm{R}}\). 
The AUC is computed by varying the detection threshold, providing a threshold-independent measure of detection performance. 
For multi-bit watermark detection evaluation, we report \bitacc, which measures the proportion of embedded watermark bits that are correctly recovered from the watermarked \dwm or generated \dadv. 
A higher \bitacc indicates a higher proportion of the watermark bits successfully recovered during watermark detection, thereby indicating a higher watermark detection capability.

To evaluate audio fidelity, we report NISQA~\cite{nisqa} and ViSQOL~\cite{chinen2020visqolv3opensource}. NISQA (Non-Intrusive Speech Quality Assessment) is a deep-learning framework that estimates perceived speech quality from an audio sample without needing an original reference signal. This matches our threat model that the data provider does not release \dorig. We also include an intrusive (full-reference) quality metric, ViSQOL (Virtual Speech Quality Objective Listener), which estimates the perceived similarity between the reference (original) and processed (watermarked) audio. For both metrics, a higher value indicates better audio quality.

\newpar{Baseline Watermarking Schemes} We compare \tool against three state-of-the-art audio watermarking schemes: AudioSeal \cite{audioseal}, WavMark \cite{chenWavMarkWatermarkingAudio2024}, and AudioMarkNet \cite{AudioMarkNet}. We compare \tool with AudioSeal and WavMark on LibriSpeech, and with AudioMarkNet on VCTK, since these are the datasets reported in their papers. These baselines use different watermarking methods. AudioSeal is an additive waveform watermark, WavMark utilizes constrained waveform perturbations, while AudioMarkNet employs spectrogram-domain message embedding. To provide a comprehensive evaluation, we assess both payload recovery and watermark detectability for all watermarking schemes. Payload recovery performance is measured using \bitacc, while watermark presence detection is evaluated using several metrics, including F1 score and AUC. AudioMarkNet is the only open-source baseline that is reported to be radioactive; we thereby evaluate radioactivity against it.

\subsection{Watermark Effectiveness, Fidelity, and Robustness Against Distortions} \label{sec:effectiveness}

\begin{table*}[th!]
\centering
\renewcommand{\arraystretch}{1.15}
\setlength{\tabcolsep}{8pt}
\caption{Robustness and fidelity evaluation of \tool against baselines under two representative distortions on LibriSpeech and VCTK. Complete results of no-distortion and all 11 evaluated distortion settings are reported in Appendix~\ref{app:distortion}.}
\label{tab:distortion_specific_results} 
\begin{tabular}{lllcccccc|cc} 
\toprule \textbf{Distortions} & \textbf{Dataset} & \textbf{Watermark} & \textbf{\detacc} & \textbf{\bitacc} & \textbf{P~(\%)} & \textbf{R~(\%)} & \textbf{F1~(\%)} & \textbf{AUC} & \textbf{ViSQOL} & \textbf{NISQA} \\ 
\midrule
\multirow{5}{*}{Phase Shift} & \multirow{3}{*}{LibriSpeech} & AudioSeal & 52.00  & 50.50 & \textbf{100.00} & 4.00   & 7.69   & 0.520 & 3.41 & 2.41 \\ 
& & WavMark & 59.00  & 15.94  & \textbf{100.00} & 18.00  & 30.51  & 0.590 & 3.42 & 2.41 \\ 
& & \tool  & \textbf{99.50} & \textbf{94.34} & 99.01 & \textbf{100.00} & \textbf{99.50} & \textbf{0.995} & 3.23 & 2.23  \\ 
\cmidrule{2-11} & \multirow{2}{*}{VCTK} & AudioMarkNet & 51.50 & 48.33 & 71.43  & 5.00  & 9.35  & 0.515 & 2.99 & 2.45 \\ 
& & \tool & \textbf{99.00}  & \textbf{87.34} & \textbf{98.04}  & \textbf{100.00} & \textbf{99.01}  & \textbf{0.990} & 2.86 & 2.41 \\ 
\midrule 
\multirow{5}{*}{Gaussian 20dB} & \multirow{3}{*}{LibriSpeech} & AudioSeal & 52.00  & 50.75 & 100.00 & 4.00   & 7.69   & 0.520 & 3.59 & 1.87  \\ 
& & WavMark & 50.50  & 0.88   & 100.00 & 1.00   & 1.98   & 0.505 & 3.61 & 1.92 \\ 
& & \tool & \textbf{100.00} & \textbf{97.53} & 100.00 & \textbf{100.00} & \textbf{100.00} & \textbf{1.000} & 3.58 & 2.47 \\ \cmidrule{2-11} 
& \multirow{2}{*}{VCTK} & AudioMarkNet & 83.50 & 80.76 & \textbf{100.00} & 67.00 & 80.24 & 0.835 & 3.33 & 1.96  \\ 
& & \tool & \textbf{99.50}  & \textbf{92.81} & 99.01  & \textbf{100.00} & \textbf{99.50}  & \textbf{0.995} & 3.33 & 2.93  \\ 
\bottomrule 
\end{tabular}
\end{table*}
In this section, we begin by evaluating \tool's watermark effectiveness and fidelity on \dwm. Table~\ref{tab:distortion_results_comparison} first compares \tool with existing watermarking baselines in the absence of distortions, and then compares robustness against various distortions.

\newpar{Effectiveness and Fidelity} 
Without distortions, \tool achieves watermark effectiveness and fidelity comparable to prior methods on both datasets. All watermarks exhibit high \bitacc except  AudioSeal, which performs only slightly better than random guessing. 
As for fidelity, \tool consistently achieves the highest NISQA scores, while obtaining lower ViSQOL than baselines. 
This is expected, as previous watermark schemes embed watermarks by adding carefully constrained perturbations to the waveform or time-frequency representation, often guided by psychoacoustic masking~\cite{chenWavMarkWatermarkingAudio2024,audioseal,li2025harmonicattackadaptivecrossdomainaudio}. Since these methods preserve the original signal semantics, full-reference metrics such as ViSQOL will naturally report high similarity to the source audio.
In contrast, \tool embeds watermarks through controlled semantic shifts in the semantic latent representation of an audio encoder; hence, the watermarked audio is not intended to be an exact waveform/acoustic replica of the original, which can lower ViSQOL despite preserved perceptual quality. This interpretation is supported by the NISQA results. Unlike ViSQOL, NISQA is a no-reference metric that evaluates audio quality without access to the original audio. The consistently higher NISQA scores indicate that \tool preserves perceptual naturalness and avoids audible artifacts despite modifying the underlying semantic representation. This is enabled by the joint training objectives (See~\S~\ref{sec:loss}), which explicitly balance watermark detectability and fidelity.

To further demonstrate that \dwm generated by \tool, despite introducing a semantic shift, does not change the content of the underlying audio, we use Whisper captioning models~\cite{whisper}. Specifically, we transcribe \dwm using five Whisper models with different sizes and compare the generated captions to the original text with character-level and word-level similarity, ranging from 0~to~1. 
On average, \tool achieves a character similarity of 0.9851 and a word similarity of 0.9522. These scores are comparable to AudioSeal (\ie~baseline with the best ViSQOL), with merely a 0.0056 drop in character similarity and a 0.0271 drop in word similarity. The high similarities are consistent across all model sizes, with mainly only punctuation differences, indicating that \tool preserves audio content. 
Detailed results are in Table~\ref{tab:captioning_similarity} in Appendix~\ref{app:caption}.

\newpar{Robustness Against Distortions} We measure whether the watermark can still be reliably detected under a comprehensive list of 11 common audio distortions (See Appendix~\ref{app:hyper} for hyperparameters).
We present the results of \tool against its baselines in Table~\ref{tab:distortion_results_comparison}. The ``Without'' rows report the setting without any distortions, \ie~only with the watermark applied. The ``With'' rows report results averaged over the ``Without'' setting and the 11 evaluated distortions, \ie~12 settings in total. The results show that \tool maintains near-perfect detection, F1, and AUC on both LibriSpeech and VCTK samples under distortions. Average BRRs of 97.94\% and 95.08\% on the two datasets indicate that bit recovery is largely unaffected by the distortions. \tool consistently outperforms all the baselines in \detacc, \bitacc, F1, and AUC on the average distortion results.

We report detailed results on two representative distortions in Table~\ref{tab:distortion_specific_results}, and provide the full results of the 11 evaluated distortions for \tool and the baselines in Appendix~\ref{app:distortion}. We can observe that none of the baselines is robust against phase shift, as they all show around 50\% \detacc and \bitacc, indicating random guessing. In contrast, \tool has 99.5\% \detacc and 94.34\% \bitacc on LibriSpeech, and 99\% \detacc and 87.34\% \bitacc on VCTK against phase shift. 
In addition, AudioSeal and WavMark are not robust against Gaussian noise. For instance, when the noise level is 20dB, AudioSeal's \detacc drops to 52\% with 50.75\% \bitacc, and WavMark's \detacc drops to 50.5\% with near-zero \bitacc, indicating the detector predicts flipped bits under Gaussian noise. These results indicate that baselines do not achieve uniformly strong robustness against all types of distortions, while simple distortions such as Gaussian noise are sufficient to largely degrade baseline watermarks' detection.

\begin{table}[th!]
\centering
\small
\setlength{\tabcolsep}{1.25pt}
\caption{Watermark effectiveness across different bit sizes.}
\label{tab:bit_size}
\begin{tabular}{lcccccc|cc}
\toprule
\textbf{\#bits} & \textbf{\detacc} & \textbf{\bitacc} & \textbf{P~(\%)} & \textbf{R~(\%)} & \textbf{F1~(\%)} & \textbf{AUC} & \textbf{ViSQOL} & \textbf{NISQA} \\
\midrule
16 & 100.00 & 99.88 & 100.00 & 100.00 & 100.00 & 1.000 & 4.52 & 4.68 \\
32 & 100.00 & 99.75 & 100.00 & 100.00 & 100.00 & 1.000 & 4.44 & 4.66 \\
48 & 100.00 & 98.67 & 100.00 & 100.00 & 100.00 & 1.000 & 4.28 & 4.06 \\
\bottomrule
\end{tabular}
\end{table}

\begin{table*}[th!]
\caption{Performance of \tool against baseline competitors under codec, optimization, and learning-based attacks. \tool is trained on LibriSpeech, so all the results on VCTK of \tool are cross-dataset transfer results.}
\centering
\renewcommand{\arraystretch}{1.2} 
\begin{tabular}{lllcccccc}
\toprule
\textbf{Attack} & \textbf{Dataset} & \textbf{Watermark} & \textbf{\detacc~(\%)} & \textbf{\bitacc~(\%)} & \textbf{P~(\%)} & \textbf{R~(\%)} & \textbf{F1~(\%)} & \textbf{AUC} \\
\midrule

\multirow{5}{*}{MP3/OGG/Opus}
  & \multirow{3}{*}{LibriSpeech}
    & WavMark   & 62.50 & 23.31 & \textbf{100.00} & 25.00 & 40.00 & 0.625 \\
  & & AudioSeal & 51.00 & 47.44 & \textbf{100.00} & 2.00  & 3.92  & 0.510 \\
  & & \tool    & \textbf{87.50} & \textbf{84.88} & \textbf{100.00} & \textbf{75.00} & \textbf{85.71} & \textbf{0.875} \\
\cmidrule{2-9}
  & \multirow{2}{*}{VCTK}
    & AudioMarkNet & 94.00 & \textbf{100.00} & \textbf{100.00} & 88.00 & 93.62 & 0.940 \\
  & & \tool        & \textbf{99.50} & 82.97  & \textbf{100.00} & \textbf{99.00} & \textbf{99.50} & \textbf{0.995} \\
\midrule

\multirow{5}{*}{EnCodec}
  & \multirow{3}{*}{LibriSpeech}
    & WavMark   & 50.00 & 0.00  & 0.00   & 0.00  & 0.00  & 0.500 \\
  & & AudioSeal & 56.50 & 48.69 & \textbf{100.00} & 13.00 & 23.01 & 0.565 \\
  & & \tool    & \textbf{85.00} & \textbf{53.84} & \textbf{100.00} & \textbf{70.00} & \textbf{82.35} & \textbf{0.850} \\
\cmidrule{2-9}
  & \multirow{2}{*}{VCTK}
    & AudioMarkNet & 50.00 & 56.50 & 0.00   & 0.00  & 0.00  & 0.500 \\
  & & \tool        & \textbf{88.50} & \textbf{69.88} & \textbf{100.00} & \textbf{77.00} & \textbf{87.01} & \textbf{0.885} \\
\midrule

\multirow{5}{*}{\audiomarkbench}
  & \multirow{3}{*}{LibriSpeech}
    & WavMark   & 52.00 & 3.75  & \textbf{100.00} & 4.00  & 7.69  & 0.520 \\
  & & AudioSeal & 61.50 & 49.75 & \textbf{100.00} & 23.00 & 37.40 & 0.615 \\
  & & \tool    & \textbf{95.50} & \textbf{94.09} & \textbf{100.00} & \textbf{91.00} & \textbf{95.29} & \textbf{0.955} \\
\cmidrule{2-9}
  & \multirow{2}{*}{VCTK}
    & AudioMarkNet & 50.00 & 72.38 & 0.00   & 0.00  & 0.00  & 0.500 \\
  & & \tool        & \textbf{92.50} & \textbf{81.41} & \textbf{100.00} & \textbf{85.00} & \textbf{91.89} & \textbf{0.925} \\
\midrule

\multirow{5}{*}{HarmonicAttack}
  & \multirow{3}{*}{LibriSpeech}
    & WavMark   & 64.50 & 23.88 & \textbf{100.00} & 29.00 & 44.96 & 0.645 \\
  & & AudioSeal & 50.00 & 51.54 & 0.00   & 0.00  & 0.00  & 0.500 \\
  & & \tool    & \textbf{98.50} & \textbf{85.97} & \textbf{100.00} & \textbf{97.00} & \textbf{98.48} & \textbf{0.985} \\
\cmidrule{2-9}
  & \multirow{2}{*}{VCTK}
    & AudioMarkNet & 50.00    & 67.69    &  0.00    & 0.00    & 0.00    & 0.500   \\
  & & \tool        & \textbf{99.50} & \textbf{72.38} & \textbf{100.00} & \textbf{99.00} & \textbf{99.50} & \textbf{0.995} \\

\bottomrule
\end{tabular}
\label{table-perf-audioseal-librispeech}
\end{table*}

\newpar{Effect of Bit Sizes}
For evaluation completeness, we also evaluate \tool's effectiveness and robustness against common distortions with varying bit sizes, specifically 16, 32, and 48 bits.
As Table~\ref{tab:bit_size} displays, \tool shows 100\% \detacc, F1, and AUC, and near-perfect \bitacc, regardless of bit sizes.  ViSQOL and NISQA fidelity metrics remain consistently high, with only graceful degradation at a bit size of 48. 
As for robustness, we collect results under different distortions in Tables~\ref{tab:16bits} and~\ref{tab:48bits} in Appendix~\ref{app:bit}. We observe a similar level of robustness as the 32-bit watermark (Table~\ref{tab:distortion_results_detail_lambda_libri}). When the bit size is 16, \tool achieves an average of 99.88\% \detacc and 99.27\% \bitacc under various distortions. When the bit size is 48, \tool also achieves near-perfect \detacc, F1, and AUC, with comparable \bitacc of 97.02\%.  
Overall, these results demonstrate that \tool scales reliably across different payload sizes, preserving near-perfect detectability and strong robustness while incurring only limited fidelity degradation at larger bit sizes.

\subsection{Pre-finetuning Robustness Against Attacks} \label{subsec:robustness}
In this section, we evaluate the robustness of \tool and baseline methods against various adversarial watermark removal attacks, \ie whether the embedded watermarks (\dwm) remain detectable immediately after watermark embedding and before downstream model finetuning. Full results for post-finetuning robustness are presented in \S~\ref{sec:radio-robustness}.
Table~\ref{table-perf-audioseal-librispeech} compares the robustness of \tool and baseline watermarks under a range of codec-based attacks, optimization-based attacks (e.g., square perturbations)~\cite{andriushchenkoSquareAttackQueryefficient2020}, and adaptive learning-based attacks such as HarmonicAttack~\cite{li2025harmonicattackadaptivecrossdomainaudio}. Overall, the results reveal a clear advantage of \tool over traditional waveform/spectrogram-based watermarking schemes. It demonstrates a consistent trend: none of the existing watermarks survives all the adversarial removal attempts, whereas \tool maintains consistently stronger robustness across all evaluated attacks.

In particular, none of the evaluated baselines are robust against codec-based perturbations such as EnCodec compression (\ie around 50\% \detacc). In contrast, \tool maintains significantly higher robustness under the same setting, achieving at least 85\% \detacc in LibriSpeech and VCTK (cross-dataset transfer results).
The gap becomes much more significant under AudioSquareAttack and HarmonicAttack. 
Our only radioactive baseline, AudioMarkNet, collapses to near-random performance under both attacks, indicating complete failure of detection and message recovery. 
AudioSeal merely reaches 61.5\% \detacc under AudioSquareAttack, and completely fails under HarmonicAttack.
Similarly, WavMark completely fails under AudioSquareAttack, with slight resistance to HarmonicAttack (merely 64.5\% \detacc and 23.88\% \bitacc).
In contrast, \tool retains strong robustness under both adversarial attacks.
For instance, on LibriSpeech, \tool achieves 95.5\% \detacc and 94.09\% \bitacc under AudioSquareAttack, and 98.5\% \detacc and 85.97\% \bitacc under HarmonicAttack, demonstrating that the watermark remains reliably recoverable under strong adversarial attacks.

This robustness can be attributed to the fundamental difference in the watermark representation used by \tool. Unlike prior methods such as WavMark and AudioSeal, which embed watermark signals directly in waveform perturbations or within psychoacoustically masked time–frequency bins, \tool introduces a semantic shift by embedding watermarks in the semantic latent space of the audio representation. As a result, the watermark is not tied to localized waveform or spectrogram-level residual structures. This has two important implications.

First, semantic watermarks significantly improve robustness against optimization-based waveform attacks such as square-noise perturbations in AudioSquareAttack. Such attacks typically operate by injecting structured or random perturbations directly into the waveform, aiming to disrupt residual watermark signals. Methods like WavMark rely on explicit waveform-level embedding, so these perturbations can directly interfere with the encoded watermark signal and degrade detection. In contrast, \tool does not rely on fragile waveform residuals, making it substantially more resistant to such optimization-based corruption.

Second, semantic watermarks also provide robustness against adaptive learning-based attacks such as HarmonicAttack, which explicitly learn to identify and suppress watermark regions in the time–frequency domain. This attack is particularly effective against methods like AudioSeal and AudioMarkNet, which rely on psychoacoustic masking to localize watermark energy within perceptually insensitive time-frequency bins. They also learn a structured spectral signal to be embedded in those masked time-frequency bins. By modeling the localization information and watermark residual structure, adaptive attacks can effectively target and remove such watermark signals. \tool does not depend on fixed masked time–frequency bins or structured spectral embedding patterns. Instead, its watermark is distributed implicitly through semantic modifications of the audio content, so it can be encoded across the entire signal rather than constrained to predictable spectral locations.

Overall, these results demonstrate that \tool differs fundamentally from conventional watermarking schemes. By encoding watermark information as semantic shifts rather than waveform- or spectrogram-level perturbations, \tool avoids the core vulnerability of existing methods: reliance on localized, statistically detectable embedding structures. This design enables robustness against waveform-level optimization attacks and adaptive learning-based attacks.

\begin{table*}[th!]
\caption{Radioactivity performance comparison between \tool and AudioMarkNet. AudioMarkNet does not generalize to LibriSpeech~\cite{AudioMarkNet}, so only VCTK results are presented. The watermark encoder--decoder of \tool is trained on LibriSpeech and evaluated on LibriSpeech and VCTK (cross-dataset).}
\centering
\renewcommand{\arraystretch}{1.15}
\setlength{\tabcolsep}{8pt}
\begin{tabular}{@{}llllcccccc@{}}
\toprule
\textbf{\dwm} & \textbf{\madv} & \textbf{FT} & \textbf{Watermark}
& \textbf{\detacc~(\%)} & \textbf{\bitacc~(\%)}
& \textbf{P~(\%)} & \textbf{R~(\%)} & \textbf{F1~(\%)} & \textbf{AUC} \\
\midrule

\multirow{10}{*}{VCTK} & \multirow{2}{*}{YourTTS} & \multirow{2}{*}{Full} & AudioMarkNet
& \textbf{100.00} & \textbf{100.00} & \textbf{100.00} & \textbf{100.00} & \textbf{100.00} & \textbf{1.000} \\
& & & \tool & 99.00 & 92.06 & \textbf{100.00} & 98.00 & 98.99 & 0.990 \\

\cmidrule{2-10}

& \multirow{4}{*}{SemanticVocoder} & \multirow{2}{*}{Full} & AudioMarkNet & 50.00 & 52.69 & 0.00 & 0.00 & 0.00 & 0.500 \\
& & & \tool & \textbf{100.00} & \textbf{97.47} & \textbf{100.00} & \textbf{100.00} & \textbf{100.00} & \textbf{1.000} \\

\cmidrule{3-10}

& & \multirow{2}{*}{LoRA} & AudioMarkNet & 50.00 & 55.31 & 0.00 & 0.00 & 0.00 & 0.500 \\
& & & \tool & \textbf{100.00} & \textbf{94.16} & \textbf{100.00} & \textbf{100.00} & \textbf{100.00} & \textbf{1.000} \\

\cmidrule{2-10}

& \multirow{4}{*}{AudioLDM2} & \multirow{2}{*}{Full} & AudioMarkNet & 50.00 & 67.88 & 0.00 & 0.00 & 0.00 & 0.500 \\
& & & \tool & \textbf{98.00} & \textbf{81.09} & \textbf{100.00} & \textbf{96.00} & \textbf{97.96} & \textbf{1.000} \\

\cmidrule{3-10}

& & \multirow{2}{*}{LoRA} & AudioMarkNet & 50.00 & 50.06 & 0.00 & 0.00 & 0.00 & 0.500 \\
& & & \tool & \textbf{91.50} & \textbf{70.31} & \textbf{100.00} & \textbf{83.00} & \textbf{90.71} & \textbf{0.997} \\

\midrule

\multirow{5}{*}{LibriSpeech} & YourTTS & Full & \tool & 99.50 & 86.56 & 100.00 & 99.00 & 99.50 & 0.995 \\

\cmidrule{2-10}

& \multirow{2}{*}{SemanticVocoder} & Full & \tool & 100.00 & 96.47 & 100.00 & 100.00 & 100.00 & 1.000 \\
& & LoRA & \tool & 100.00 & 96.31 & 100.00 & 100.00 & 100.00 & 1.000 \\

\cmidrule{2-10}

& \multirow{2}{*}{AudioLDM2} & Full & \tool & 94.50 & 82.38 & 100.00 & 89.00 & 94.17 & 0.999 \\
& & LoRA & \tool & 84.50 & 73.66 & 100.00 & 69.00 & 81.66 & 0.985 \\

\bottomrule
\end{tabular}
\label{table-downstream-models}
\end{table*}

\begin{table}[th!]
\centering
\setlength{\tabcolsep}{0.6pt}
\caption{Effectiveness of size of \dwm---\dwmsize, evaluated on LibriSpeech when the full finetuning script is used. }
\label{tab:dwm_size}
\begin{tabular}{llcccccc}
\toprule
\textbf{\textbf{\madv}} & \textbf{\dwmsize} & \textbf{\detacc~(\%)} & \textbf{\bitacc~(\%)} & \textbf{P~(\%)} & \textbf{R~(\%)} & \textbf{F1~(\%)} & \textbf{AUC} \\
\midrule
\multirow{4}{*}{YourTTS}  & 100 & 99.50 & 85.34 & 100.00 & 99.00 & 99.50 & 0.995 \\
 & 500 & 99.00 & 84.56 & 100.00 & 98.00 & 98.99 & 0.990 \\
& 1,000 & 98.00 & 83.72 & 100.00 & 96.00 & 97.96 & 0.980 \\
& 2,500 & 99.50 & 86.56 & 100.00 & 99.00 & 99.50 & 0.995 \\
\midrule
\multirow{4}{*}{SemanticVocoder} & 100 & 100.00 & 98.22 & 100.00 & 100.00 & 100.00 & 1.000 \\
& 500 & 100.00 & 98.16 & 100.00 & 100.00 & 100.00 & 1.000 \\
& 1,000 & 100.00 & 95.22 & 100.00 & 100.00 & 100.00 & 1.000 \\
& 2,500 & 100.00 & 96.47 & 100.00 & 100.00 & 100.00 & 1.000 \\
\midrule
\multirow{4}{*}{AudioLDM2}  & 100 & 79.50 & 69.69 & 100.00 & 59.00 & 74.21 & 0.992 \\
 & 500 & 98.00 & 75.91 & 100.00 & 96.00 & 97.96 & 0.996 \\
& 1,000 & 95.00 & 76.34 & 100.00 & 90.00 & 94.74 & 0.995 \\
& 2,500 & 94.50 & 82.38 & 100.00 & 89.00 & 94.17 & 0.999 \\

\bottomrule
\end{tabular}
\end{table}

\subsection{Radioactivity}\label{subsec:radioactivity}
To demonstrate \tool's radioactivity, \ie~surviving downstream finetuning, we finetune three audio generation models: YourTTS, SemanticVocoder, and AudioLDM2 using LibriSpeech and VCTK as \dwm. We include VCTK as a cross-dataset evaluation. Note that AudioMarkNet~\cite{AudioMarkNet} reports its key limitation, which is that it cannot generalize to unseen speakers. In other words, they only support speakers for whom their watermark model is trained, which are those in VCTK. Therefore, we can only evaluate their non-transfer radioactivity results on VCTK. In contrast, our watermark encoder--decoder is trained only on LibriSpeech, so transfer evaluation on VCTK tests whether \tool generalizes to audio outside the training distribution.

We assume the adversary finetunes \madv with both \emph{full finetuning} and \emph{LoRA finetuning}. Full finetuning on AudioLDM2 means updating the UNet denoiser weights, and full finetuning on SemanticVocoder means updating all the trainable weights of the model. We assume the adversary finetunes \madv on 2,500 \dwm with a 32-bit watermark embedded and generates \dadv with 100 samples. For fair comparison with AudioMarkNet~\cite{AudioMarkNet}, which originally tests only on YourTTS, we use a full finetuning script equivalent to the one available in their source code.
To ensure reliable evaluation, we also test \tool on clean downstream models, \ie without finetuning on \dwm, using the same test prompts to confirm that \tool is not biased and can correctly classify negative samples.

Table~\ref{table-downstream-models} demonstrates that \tool consistently achieves strong radioactivity across all evaluated downstream generative models, finetuning settings, and \dwm. For both VCTK and LibriSpeech, \tool maintains uniformly high detection accuracy and F1 scores across all three models. For example, on VCTK, \tool achieves near-perfect radioactivity on SemanticVocoder and YourTTS with detection accuracies up to 100\%, and F1 scores up to 100\%, while still maintaining strong performance on the more challenging AudioLDM2 setting with 98\% \detacc and 97.96\% F1 under full finetuning, and 91.50\% \detacc with 90.71\% F1 under LoRA. Similar trends are observed on LibriSpeech. This indicates that \tool does not rely on a specific training corpus and can induce transferable radioactivity using arbitrary watermarked datasets \dwm. In contrast, AudioMarkNet only demonstrates effective radioactivity on the YourTTS model trained on VCTK, where it achieves perfect performance, but completely fails to generalize to the other downstream models or LibriSpeech. Specifically, AudioMarkNet collapses to random-guess performance on SemanticVocoder and AudioLDM2 with both full and LoRA finetuning, with 50\% \detacc, 0\% P/R/F1, and 0.5 AUC.

\begin{table*}[th!]
\centering
\renewcommand{\arraystretch}{1.15}
\setlength{\tabcolsep}{8pt}
\caption{Robustness of watermarks in adversarial audio \dadv generated by YourTTS after full finetuning on watermarked VCTK data (\dwm), comparing AudioMarkNet and \tool.
AudioMarkNet only supports YourTTS with full finetuning, whereas \tool supports all evaluated generation and finetuning settings.}
\label{tab:robustness_of_radioactivity_vctk_yourtts_compare}
\begin{tabular}{llcccccc}
\toprule
\textbf{Attack} & \textbf{Watermark} 
& \textbf{\detacc~(\%)} & \textbf{\bitacc~(\%)}
& \textbf{P~(\%)} & \textbf{R~(\%)} & \textbf{F1~(\%)} & \textbf{AUC} \\
\midrule

\multirow{2}{*}{Signal-level Distortions}
& AudioMarkNet  & 95.62 & 86.63 & 99.62 & 91.31 & 93.48 & 0.956 \\
& \tool         & \textbf{96.63} & \textbf{87.23} & \textbf{100.00} & \textbf{93.25} & \textbf{96.51} & \textbf{0.966} \\

\midrule

\multirow{2}{*}{\audiomarkbench}
& AudioMarkNet  & 50.00 & 61.94 & 0.00 & 0.00 & 0.00 & 0.500 \\
& \tool         & \textbf{99.50} & \textbf{72.63} & \textbf{100.00} & \textbf{99.00} & \textbf{99.50} & \textbf{0.995} \\

\midrule

\multirow{2}{*}{HarmonicAttack}
& AudioMarkNet  & 50.00 & 64.13 & 0.00 & 0.00 & 0.00 & 0.500 \\
& \tool         & \textbf{100.00} & \textbf{71.03} & \textbf{100.00} & \textbf{100.00} & \textbf{100.00} & \textbf{1.000} \\

\bottomrule
\end{tabular}
\end{table*}

\begin{table*}[th!]
\caption{\tool's robustness on \dadv generated by \madv after finetuning on watermarked VCTK data (\dwm).}
\centering
\renewcommand{\arraystretch}{1.15}
\setlength{\tabcolsep}{8pt}
\begin{tabular}{lllcccccc}
\toprule
\textbf{Attack} & \textbf{\madv} & \textbf{FT}
& \textbf{\detacc~(\%)} & \textbf{\bitacc~(\%)}
& \textbf{P~(\%)} & \textbf{R~(\%)} & \textbf{F1~(\%)} & \textbf{AUC} \\
\midrule

\multirow{5}{*}{Signal-level Distortions}
& YourTTS & Full & 96.63 & 87.23 & 100.00 & 93.25 & 96.51 & 0.966 \\\cmidrule{2-9}
& \multirow{2}{*}{SemanticVocoder} & Full & 99.96 & 93.47 & 99.92 & 100.00 & 99.96 & 1.000 \\
& & LoRA & 99.92 & 90.55 & 99.92 & 99.92 & 99.92 & 1.000 \\\cmidrule{2-9}
& \multirow{2}{*}{AudioLDM2} & Full & 98.29 & 82.53 & 99.92 & 96.67 & 98.25 & 1.000 \\
& & LoRA & 90.50 & 72.70 & 99.90 & 81.08 & 89.13 & 0.997 \\
\midrule

\multirow{5}{*}{\audiomarkbench}
& YourTTS & Full & 99.50 & 72.63 & 100.00 & 99.00 & 99.50 & 0.995 \\\cmidrule{2-9}
& \multirow{2}{*}{SemanticVocoder} & Full & 99.50 & 80.50 & 100.00 & 99.00 & 99.50 & 0.995 \\
& & LoRA & 99.00 & 77.22 & 100.00 & 98.00 & 98.99 & 0.990 \\\cmidrule{2-9}
& \multirow{2}{*}{AudioLDM2} & Full & 98.00 & 81.06 & 100.00 & 96.00 & 97.96 & 0.980 \\
& & LoRA & 90.00 & 69.53 & 100.00 & 80.00 & 88.89 & 0.900 \\

\midrule

\multirow{5}{*}{HarmonicAttack}
& YourTTS & Full & 100.00 & 71.03 & 100.00 & 100.00 & 100.00 & 1.000 \\\cmidrule{2-9}
& \multirow{2}{*}{SemanticVocoder} & Full & 99.00 & 77.50 & 98.04 & 100.00 & 99.01 & 1.000 \\
& & LoRA & 99.00 & 70.34 & 98.04 & 100.00 & 99.01 & 1.000 \\\cmidrule{2-9}
& \multirow{2}{*}{AudioLDM2} & Full & 99.00 & 72.88 & 98.04 & 100.00 & 99.01 & 0.998 \\
& & LoRA & 85.50 & 70.09 & 97.33 & 73.00 & 83.43 & 0.976 \\

\bottomrule
\end{tabular}
\label{tab:robustness_of_radioactivity_vctk}
\end{table*}

The strong and transferable radioactivity indicates that \tool's watermarks are not merely at the waveform level, which can easily be diluted or erased in finetuning, but are embedded as a semantic trait of \dwm for \madv to be effectively learned.
As expected, radioactivity on SemanticVocoder is stronger than on AudioLDM2. This is because \tool is a semantic watermark: it is embedded in high-level semantic representations. SemanticVocoder explicitly conditions generation on semantic representations, allowing the embedded watermark signal to propagate more directly through the downstream generation process and resulting in near-perfect radioactivity performance. In contrast, AudioLDM2 generates audio through an iterative latent diffusion process with a denoising UNet, which can reshape acoustic details more aggressively during generation and may partially transform the watermark signal, leading to comparatively lower but still strong radioactivity.

\newpar{Effect of the Size of \dwm} 
Table~\ref{tab:dwm_size} studies the effectiveness of the size of \dwm---\dwmsize, on \tool's radioactivity to \madv. Across all three downstream models and all evaluated \dwmsize values, \tool achieves perfect precision, indicating that every sample detected as radioactive is truly radioactive, with no false positives.
YourTTS and SemanticVocoder show similarly stable trends as \dwmsize varies. For YourTTS, \tool maintains near-perfect detection performance and high AUC across all evaluated sizes. SemanticVocoder shows an even stronger trend, achieving perfect detection metrics across all \dwmsize values while maintaining high \bitacc, indicating that it can learn and propagate \tool's watermark even from a small amount of watermarked data.
AudioLDM2 is more sensitive to \dwmsize. In the most challenging setting where \dwmsize is only 100, the detector becomes conservative, leading to lower recall despite perfect precision and a high AUC of 0.992. 
However, performance improves quickly as \dwmsize increases: with 500 watermarked samples, recall already reaches 96\%. Further increase in \dwmsize yields further \bitacc improvements.

\subsection{Post-finetuning Robust Radioactivity}
\label{sec:radio-robustness}
Just as various attack attempts on the watermarked samples (\S~\ref{subsec:robustness}), an active attacker may seek to remove the watermarks on the generated samples \dadv from the finetuned downstream generation models \madv. Therefore, evaluating the robustness of watermarks in \dadv is also important. 

As reported in Table \ref{table-downstream-models}, AudioMarkNet is only radioactive on VCTK samples from YourTTS. Therefore, we only compare watermark robustness in \dadv (VCTK) against AudioMarkNet on YourTTS (Table~\ref{tab:robustness_of_radioactivity_vctk_yourtts_compare}) while reporting our robustness also on the other two models and LibriSpeech (Table~\ref{tab:robustness_of_radioactivity_vctk} and Table \ref{tab:robustness_of_radioactivity_libri}). As shown in Table \ref{tab:robustness_of_radioactivity_vctk_yourtts_compare}, \tool consistently demonstrates stronger robust radioactivity than AudioMarkNet across all attacks. Under common signal-level distortions, \tool excels on all evaluated metrics compared to AudioMarkNet on YourTTS. More importantly, under stronger attacks such as \audiomarkbench and HarmonicAttack, AudioMarkNet completely collapses to random-guess behavior with 50\% \detacc, and 0\% precision/recall/F1, whereas \tool maintains near-perfect robustness with up to 100\% \detacc and F1 score.

Table \ref{tab:robustness_of_radioactivity_vctk} and Table~\ref{tab:robustness_of_radioactivity_libri}~(Appendix~\ref{app:robustness_of_radioactivity}) further demonstrate that this robustness generalizes across different downstream generation models, finetuning strategies, and \dwm. In most settings, \tool achieves above 90\% of \detacc and F1 score. Even on the most challenging AudioLDM2 LoRA-tuned model, \tool is still robust to the strongest HarmonicAttack, with \detacc at 85.50\% and AUC at 0.976. The results are comparable to the pre-finetuning robustness results on the directly watermarked samples \dwm presented in Table~\ref{tab:distortion_results_comparison} and Table~\ref{table-perf-audioseal-librispeech}. This demonstrates that \tool is a semantic watermark whose modification to the audio involves a subtle yet robust semantic footprint that successfully carries through various downstream model adaptations, which is a property that none of the existing watermarks achieve.

\begin{figure*}[th!]
    \centering

    \begin{subfigure}[t]{0.32\textwidth}
        \centering
        \includegraphics[width=\linewidth]{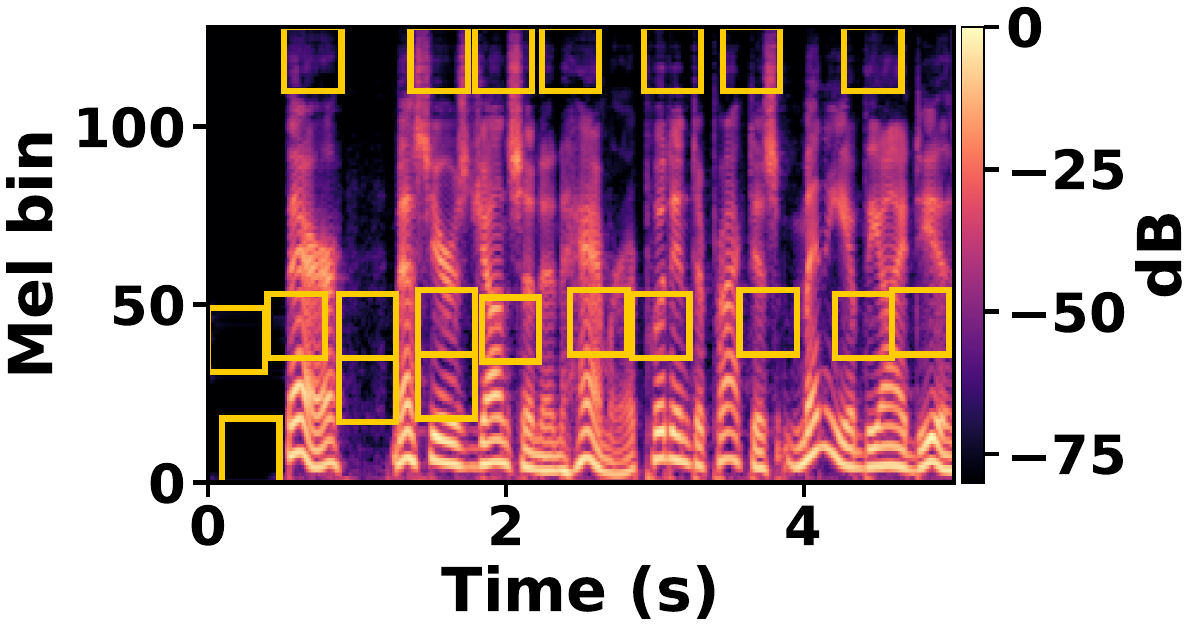}
        \caption{AudioSeal Sample 1.}
    \end{subfigure}
    \hfill
    \begin{subfigure}[t]{0.32\textwidth}
        \centering
        \includegraphics[width=\linewidth]{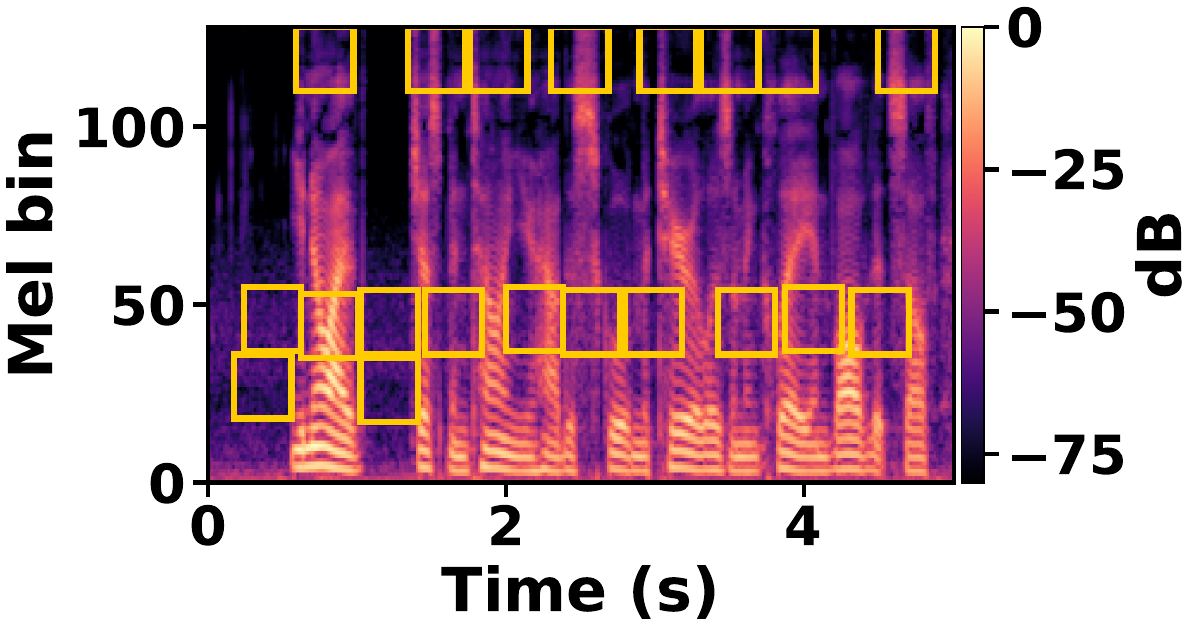}
        \caption{AudioSeal Sample 2.}
    \end{subfigure}
    \hfill
    \begin{subfigure}[t]{0.32\textwidth}
        \centering
        \includegraphics[width=\linewidth]{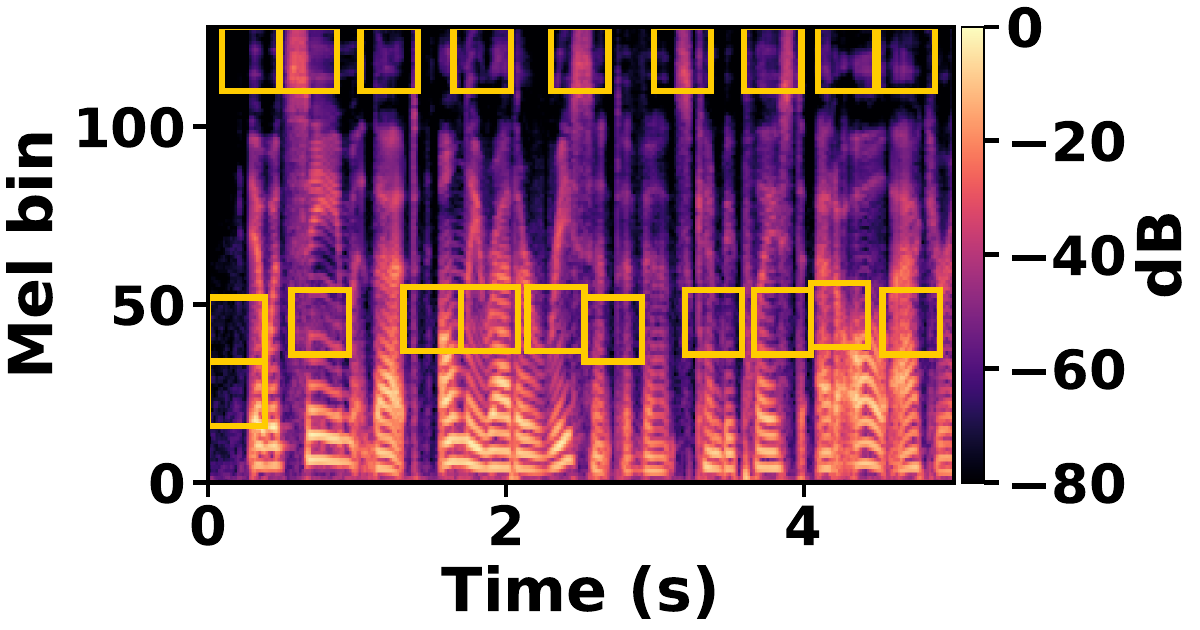}
        \caption{AudioSeal Sample 3.}
    \end{subfigure}

    \vspace{0.5em}

    \begin{subfigure}[t]{0.32\textwidth}
        \centering
        \includegraphics[width=\linewidth]{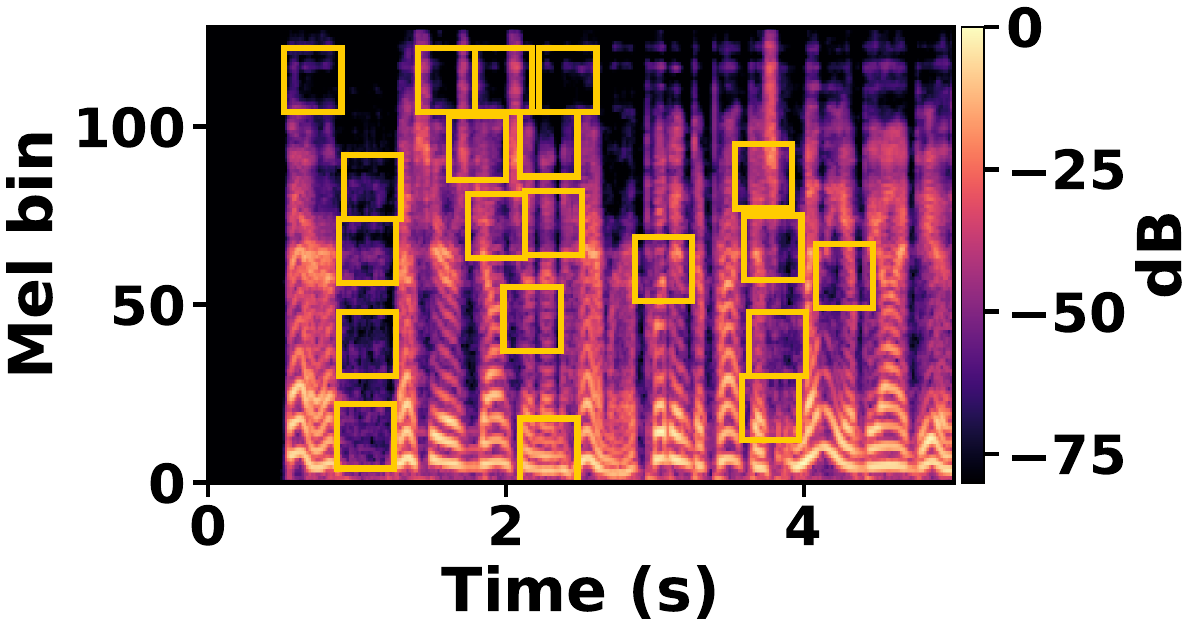}
        \caption{\tool Sample 1.}
    \end{subfigure}
    \hfill
    \begin{subfigure}[t]{0.32\textwidth}
        \centering
        \includegraphics[width=\linewidth]{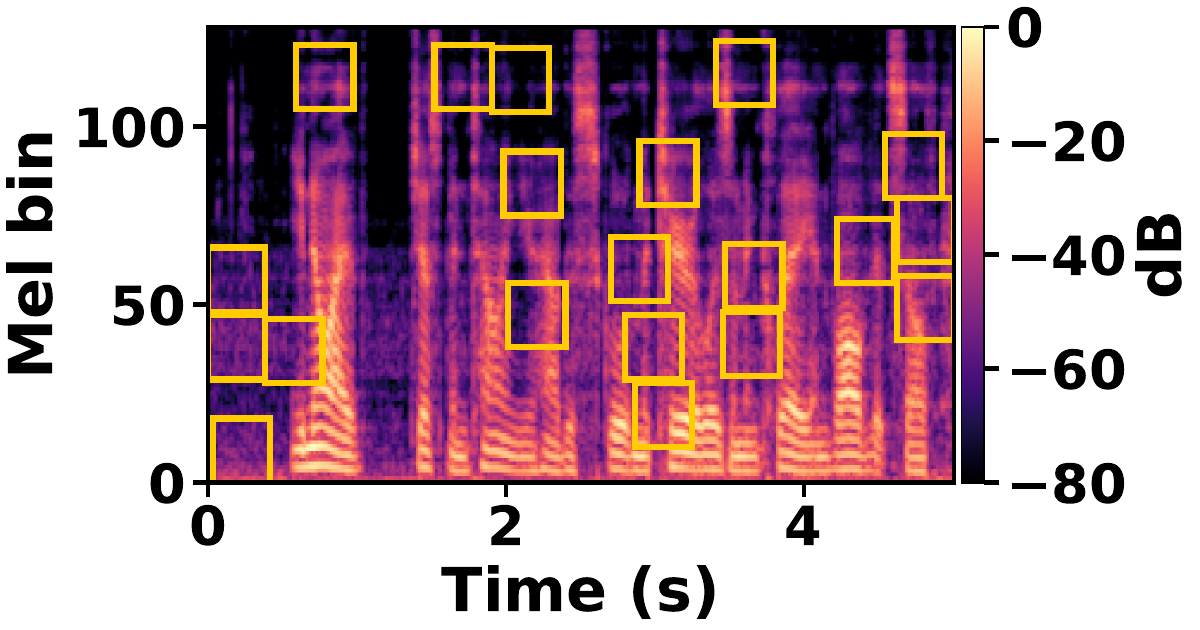}
        \caption{\tool Sample 2.}
    \end{subfigure}
    \hfill
    \begin{subfigure}[t]{0.32\textwidth}
        \centering
        \includegraphics[width=\linewidth]{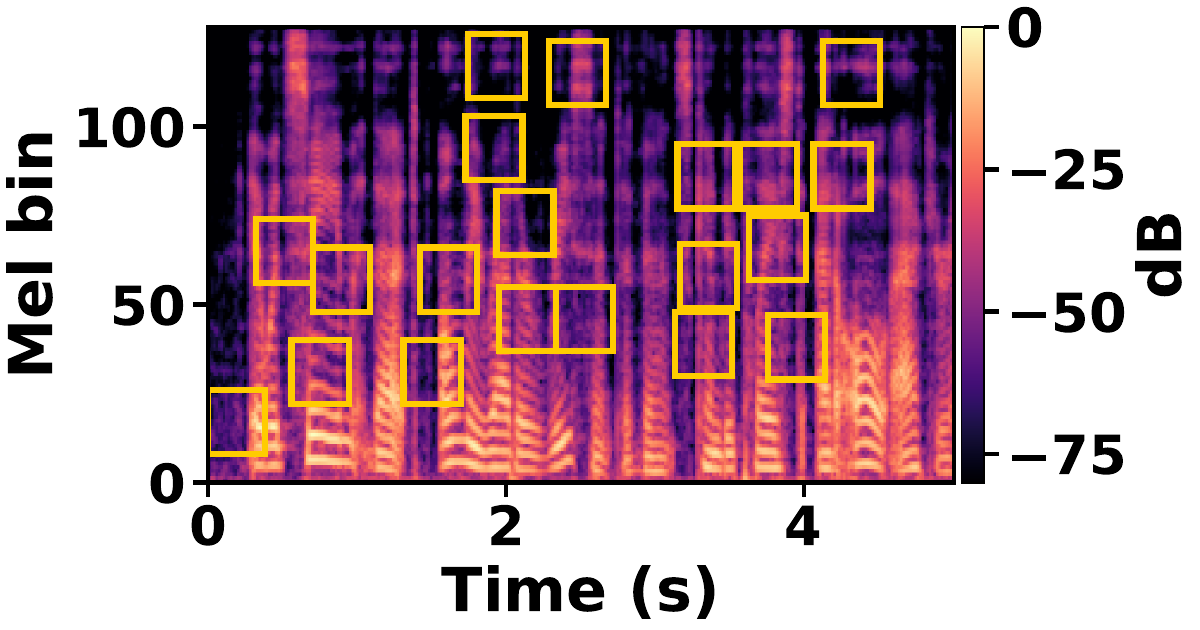}
        \caption{\tool Sample 3.}
    \end{subfigure}

    \caption{Mel spectrogram visualizations of watermarked audio produced by \tool and AudioSeal across three representative LibriSpeech samples. 
    Yellow boxes indicate the strongest residual bins in mel space (top-$20$).
    }
    \label{fig:watermarked_spectrograms_mel}
\end{figure*}

\subsection{Spectrogram Analysis} \label{subsec:spectrogram}
Figures~\ref{fig:watermarked_spectrograms_mel} and~\ref{fig:watermarked_spectrograms_stft} (Appendix~\ref{app:spec}) visualize the watermarked audio produced by \tool and AudioSeal~\cite{audioseal}, using both mel spectrograms and STFT spectrograms. The mel spectrograms represent spectral energy over perceptually motivated mel-frequency bins, emphasizing perceptually meaningful frequency bands. It spans 128 mel-frequency bins, with lower bin indices corresponding to lower frequencies. In contrast, STFT spectrograms show spectral energy over linearly spaced frequency bins.
For each watermarked sample, we compare it with the corresponding original audio and highlight the time-frequency bins with the strongest residual energy. These highlighted bins provide an approximate visualization of where the watermark signal is concentrated.
For better visualization, we select the top-$K$ ($K=20$) residual bins in mel space and linear-frequency STFT space. 

From both spectrograms, the watermarked regions produced by \tool are distributed across multiple time-frequency bins, whereas AudioSeal appears more constrained by psychoacoustic masking effects. This is particularly obvious in mel spectrograms (Figure~\ref{fig:watermarked_spectrograms_mel}), where AudioSeal's residual regions are concentrated around localized high-energy speech components, while \tool spreads the watermark perturbation across broader perceptually relevant mel-frequency bands. This suggests that \tool embeds the watermark in a more distributed and content-aware manner, rather than relying primarily on localized masked regions.
\tool's sparse and globalized watermark embedding aligns with the goal of semantic watermarking: the watermark artifacts should be tied to the semantic structures of the audio rather than low-level spectral perturbations. 

Overall, the spectrogram analysis provides intuitive explanations for \tool's robustness to common distortions, adversarial attacks, and its radioactive effect on downstream generative models: the embedded watermark is not merely attached as an acoustic artifact, but is entangled with the semantic structure that downstream models can learn from the watermark-shifted semantic distribution. 

\section{Limitations, Future Work, and Discussion}\label{sec:discuss}

For audio generation models that have continuous acoustic, semantic, or speaker-conditioned representations, such as AudioLDM2,  SemanticVocoder, and YourTTS, a small, consistent semantic shift across all watermarked samples introduced by \tool can be learned more easily during finetuning. In contrast, token-based speech generation pipelines may partially disrupt these signals, since quantization can discard or reshape fine-grained spectral or semantic patterns that were used to encode the watermark. In other words, \tool is less radioactive to downstream models that rely on discrete speech tokenization.

We observe this limitation on Spark-TTS~\cite{wang2025sparkttsefficientllmbasedtexttospeech}, an LLM-based text-to-speech model with discrete speech representations. 
Compared with the strong radioactivity observed on AudioLDM2 and SemanticVocoder in Section~\ref{subsec:radioactivity}, \tool transfers less effectively to Spark-TTS, reaching a bit recovery rate of 74.38\% with increased watermark strength. Although this result is notably lower than those obtained on the other downstream models, it remains clearly above random guessing for a 32-bit watermark, indicating that the watermark signal is still partially preserved even through discrete speech tokenization. However, increasing watermark strength can compromise perceptual quality, introducing a trade-off between downstream traceability and audio fidelity.

This suggests an important direction for future work: improving radioactivity for token-based speech generation models may require watermark designs that are explicitly aligned with the model's tokenization space. More broadly, this motivates the development of semantic watermarks that remain robust not only under continuous generative modeling but also under LM-based discrete speech tokenization. 

Beyond robust radioactivity, \tool supports the responsible development of semantic audio generation systems.
Semantic generation offers socially valuable capabilities by making audio synthesis more controllable, interpretable, and aligned with human-level concepts such as speech content, speaker identity, emotion, and musical structure.
Such capabilities support socially beneficial applications, \eg safer audio editing, accessible creative tools, and more interpretable human--AI interaction.
As such models become increasingly realistic and widely deployed~\cite{xiaomidasheng,xie2026semanticvocoderbridgingaudiogeneration,dinkel2026midashenglmefficientaudiounderstanding}, watermarking schemes should also operate at the semantic level. \tool takes a step in this direction by embedding traceable signals into the audio's semantic traits.

\section{Related Work}\label{sec:rw}
\newpar{Audio Watermarking}
Existing methods can be divided into non-radioactive and radioactive watermarking. Non-radioactive methods
watermark audio samples after generation. 
AudioSeal~\cite{audioseal} is zero-bit; it jointly trains a generator-detector pair that embeds waveform perturbations optimized under psychoacoustic masking principles.
WavMark~\cite{chenWavMarkWatermarkingAudio2024} is multi-bit; it uses invertible neural networks on STFT representations to embed and decode watermark bits, while Timbre Watermarking~\cite{liuDetectingVoiceCloning2023} embeds repeated watermark patterns into speaker spectral characteristics for robustness.

Radioactive watermarking methods aim to make downstream generative models inherit the watermark. For instance, AudioMarkNet~\cite{AudioMarkNet} tailors its watermarks to 11 speakers and the YourTTS model, so that the downstream YourTTS model reproduces the watermark for these speakers after adaptation. However, AudioMarkNet's watermarks are not generalizable to unseen speakers and other models.
Latent Watermarking~\cite{romanLatentWatermarkingAudio2025} has a different objective from other existing watermarks: model provenance. Despite its name, and although their watermarks propagate through codec-derived latents, they also embed watermarks at the waveform level. Specifically, it watermarks training waveforms with an AudioSeal-style signal robust to EnCodec tokenization, then trains a MusicGen LM on the resulting tokens.
We do not include comparisons with Latent Watermarking~\cite{romanLatentWatermarkingAudio2025} because the authors have not publicly released their source code.
Regardless of architectural differences, most existing methods embed structured perturbations into waveform or spectrogram space, relying on psychoacoustic masking and low-level residual patterns rather than semantic latent variations, making them vulnerable to adversarial attacks. 

\newpar{Audio Watermark Removal}
Existing audio watermark removal attacks can be broadly categorized into signal-distortion-based, optimization-based, and adaptive learning-based approaches. Signal-distortion-based attacks \cite{oreillyDeepAudioWatermarks2025,ozerComprehensiveRealWorldAssessment2025} apply fixed transformations such as compression, filtering, resampling, or additive noise to weaken watermark signals. Optimization-based attacks instead search for sample-specific perturbations that minimize detector confidence while preserving perceptual quality, including query-based and adversarial search methods such as square attacks~\cite{liuAudioMarkBenchBenchmarkingRobustness}. More recently, adaptive learning-based attacks explicitly model the structure of watermark residuals. HarmonicAttack~\cite{li2025harmonicattackadaptivecrossdomainaudio} shows that psychoacoustically masked watermark perturbations often exhibit learnable spectral placement and residual patterns that can be estimated and removed through adaptive optimization, exposing the vulnerability of waveform- and spectrogram-level watermarking schemes.

\section{Conclusion}\label{sec:conclusion}
In this paper, we introduced \tool, a radioactive multi-bit semantic audio watermarking scheme for protecting audio against unauthorized use in downstream generative models. Unlike prior audio watermarking methods that embed signals as waveform- or spectrogram-level perturbations, \tool injects multi-bit perturbations into semantic audio latent representations. This design makes the watermark semantically aligned, allowing it to maintain consistently high detectability not only after common audio manipulations and adversarial removal attacks, but also after downstream model finetuning and subsequent adversarial removals.
It achieves robust and reliable detection and bit recovery under diverse signal-level distortions and downstream finetuning settings, which enables voice fraud detection and black-box ownership verification for protected audio data. 
Overall, \tool suggests that semantic-latent-based transformation is a promising direction for audio watermarking.

\clearpage

\bibliographystyle{IEEEtran}
\bibliography{ref}

@misc{chinen2020visqolv3opensource,
      title={{ViSQOL v3: An Open Source Production Ready Objective Speech and Audio Metric}}, 
      author={Michael Chinen and Felicia S. C. Lim and Jan Skoglund and Nikita Gureev and Feargus O'Gorman and Andrew Hines},
      year={2020},
      eprint={2004.09584},
      archivePrefix={arXiv},
      primaryClass={eess.AS},
      url={https://arxiv.org/abs/2004.09584}, 
}

@inproceedings{nisqa,
  title = {{NISQA: A Deep CNN-Self-Attention Model for Multidimensional Speech Quality Prediction with Crowdsourced Datasets}},
  author = {Mittag, Gabriel and Naderi, Babak and Chehadi, Assmaa and M{\"o}ller, Sebastian},
  booktitle = {Interspeech 2021},
  pages = {2127--2131},
  year = {2021},
  month = aug,
  publisher = {ISCA},
  doi = {10.21437/Interspeech.2021-299},
  url = {http://dx.doi.org/10.21437/Interspeech.2021-299}
}

@misc{li2025harmonicattackadaptivecrossdomainaudio,
      title={{HarmonicAttack: An Adaptive Cross-Domain Audio Watermark Removal}}, 
      author={Kexin Li and Xiao Hu and Ilya Grishchenko and David Lie},
      year={2025},
      eprint={2511.21577},
      archivePrefix={arXiv},
      primaryClass={cs.SD},
      url={https://arxiv.org/abs/2511.21577}, 
}

@misc{xiaomidasheng,
      title={{Scaling up masked audio encoder learning for general audio classification}}, 
      author={Heinrich Dinkel and Zhiyong Yan and Yongqing Wang and Junbo Zhang and Yujun Wang and Bin Wang},
      year={2024},
      eprint={2406.06992},
      archivePrefix={arXiv},
      primaryClass={cs.SD},
      url={https://arxiv.org/abs/2406.06992}, 
}

@misc{xie2026semanticvocoderbridgingaudiogeneration,
      title={{SemanticVocoder: Bridging Audio Generation and Audio Understanding via Semantic Latents}}, 
      author={Zeyu Xie and Chenxing Li and Qiao Jin and Xuenan Xu and Guanrou Yang and Wenfu Wang and Mengyue Wu and Dong Yu and Yuexian Zou},
      year={2026},
      eprint={2602.23333},
      archivePrefix={arXiv},
      primaryClass={cs.SD},
      url={https://arxiv.org/abs/2602.23333}, 
}

@misc{dinkel2026midashenglmefficientaudiounderstanding,
      title={{MiDashengLM: Efficient Audio Understanding with General Audio Captions}}, 
      author={Heinrich Dinkel and Gang Li and Jizhong Liu and Jian Luan and Yadong Niu and Xingwei Sun and Tianzi Wang and Qiyang Xiao and Junbo Zhang and Jiahao Zhou},
      year={2026},
      eprint={2508.03983},
      archivePrefix={arXiv},
      primaryClass={cs.SD},
      url={https://arxiv.org/abs/2508.03983}, 
}

@misc{wen2023treeringwatermarksfingerprintsdiffusion,
      title={{Tree-Ring Watermarks: Fingerprints for Diffusion Images that are Invisible and Robust}}, 
      author={Yuxin Wen and John Kirchenbauer and Jonas Geiping and Tom Goldstein},
      year={2023},
      eprint={2305.20030},
      archivePrefix={arXiv},
      primaryClass={cs.LG},
      url={https://arxiv.org/abs/2305.20030}, 
}

@misc{ci2024ringidrethinkingtreeringwatermarking,
      title={{RingID: Rethinking Tree-Ring Watermarking for Enhanced Multi-Key Identification}}, 
      author={Hai Ci and Pei Yang and Yiren Song and Mike Zheng Shou},
      year={2024},
      eprint={2404.14055},
      archivePrefix={arXiv},
      primaryClass={cs.CV},
      url={https://arxiv.org/abs/2404.14055}, 
}

@misc{haas2024discoveringinterpretabledirectionssemantic,
      title={{Discovering Interpretable Directions in the Semantic Latent Space of Diffusion Models}}, 
      author={René Haas and Inbar Huberman-Spiegelglas and Rotem Mulayoff and Stella Graßhof and Sami S. Brandt and Tomer Michaeli},
      year={2024},
      eprint={2303.11073},
      archivePrefix={arXiv},
      primaryClass={cs.CV},
      url={https://arxiv.org/abs/2303.11073}, 
}

@misc{li2025hmarkradioactivemultibitsemanticlatent,
      title={{HMARK: Radioactive Multi-Bit Semantic-Latent Watermarking for Diffusion Models}}, 
      author={Kexin Li and Guozhen Ding and Ilya Grishchenko and David Lie},
      year={2025},
      eprint={2512.00094},
      archivePrefix={arXiv},
      primaryClass={cs.CR},
      url={https://arxiv.org/abs/2512.00094}, 
}

@misc{AP_2024_BidenRobocallLawsuit,
  author       = {{Holly Ramer}},
  publisher = {Associated Press},
  title        = {{The League of Women Voters is suing those involved in robocalls sent to New Hampshire voters}},
  year         = {2024},
  month        = mar,
  day          = {14},
  howpublished = {\url{https://apnews.com/article/new-hampshire-primary-biden-robocalls-f7cc3c2610ab6ccdfaa26d2a8ea1cbdb}},
}

@misc{LWV_2025_KramerDefaultJudgment,
  author       = {{League of Women Voters}},
  title        = {{League of Women Voters and Free Speech For People Applaud Judgment for Plaintiffs in New Hampshire Voter Intimidation Lawsuit}},
  year         = {2025},
  month        = nov,
  day          = {24},
  howpublished = {\url{https://www.lwv.org/newsroom/press-releases/league-women-voters-and-free-speech-people-applaud-judgment-plaintiffs-new}},
}

@misc{Reuters_2025_LovoVoiceActors,
  author       = {Brittain, Blake},
  title        = {{Voice actors can pursue some claims over AI voiceovers, US court says}},
  year         = {2025},
  month        = jul,
  day          = {10},
  howpublished = {\url{https://www.reuters.com/legal/litigation/voice-actors-can-pursue-some-claims-over-ai-voiceovers-us-court-says-2025-07-10/}},
}

@misc{Reuters_2024_SunoUdioLawsuits,
  author       = {Brittain, Blake},
  title        = {{Music AI startups Suno and Udio slam record label lawsuits in court filings}},
  year         = {2024},
  month        = aug,
  day          = {1},
  howpublished = {\url{https://www.reuters.com/legal/litigation/music-ai-startups-suno-udio-slam-record-label-lawsuits-court-filings-2024-08-01/}},
}

@misc{Reuters_2025_UdioSettlement,
  author       = {{Rishabh Jaiswal}},
  publisher = {Reuters},
  title        = {{Universal Music settles copyright dispute with AI firm Udio}},
  year         = {2025},
  month        = oct,
  day          = {29},
  howpublished = {\url{https://www.reuters.com/business/media-telecom/universal-music-settles-copyright-dispute-with-ai-firm-udio-2025-10-30/}},
}

@misc{Reuters_2025_WarnerSunoSettlement,
  author       = {{Reuters}},
  title        = {{Warner Music Group settles copyright case with Suno for licensed AI music}},
  year         = {2025},
  month        = nov,
  day          = {25},
  howpublished = {\url{https://www.reuters.com/legal/litigation/warner-music-group-settles-copyright-case-with-suno-licensed-ai-music-2025-11-25/}},
}

@misc{Justia_2026_UdioMotionDismiss,
  author       = {{United States District Court for the Southern District of New York}},
  title        = {{UMG Recordings, Inc. et al. v. Uncharted Labs, Inc. et al., No. 1:2024cv04777, Document 156}},
  year         = {2026},
  month        = apr,
  day          = {15},
  howpublished = {\url{https://law.justia.com/cases/federal/district-courts/new-york/nysdce/1%3A2024cv04777/623701/156/}},
  note         = {Order and Opinion Denying Defendants' Motion to Dismiss}
}

@misc{audioseal,
      title={{Proactive Detection of Voice Cloning with Localized Watermarking}}, 
      author={Robin San Roman and Pierre Fernandez and Alexandre Défossez and Teddy Furon and Tuan Tran and Hady Elsahar},
      year={2024},
      eprint={2401.17264},
      archivePrefix={arXiv},
      primaryClass={cs.SD},
      url={https://arxiv.org/abs/2401.17264}, 
}

@mastersthesis{kanellopoulou2025unauthorized,
author = {Kanellopoulou, Christina},
title = {{Unauthorized Voice Cloning: The Legal Response in the Intersection of Performers' Rights, Sound Recording Protection, and Image Rights in the Age of AI}},
type = {Dissertation},
school = {Stockholm University},
year = {2025}
}

@article{iapp2023voiceactors,
  author       = {Chuks-Okeke, Ekene and Adetunji, Ade and Leong, Brenda},
  title        = {{Voice actors and generative AI: Legal challenges and emerging protections}},
  journal      = {IAPP News},
  year         = {2023},
  month        = {Dec},
  url          = {https://iapp.org/news/a/voice-actors-and-generative-ai-legal-challenges-and-emerging-protections}


}

@misc{kimiteamKimiAudioTechnicalReport2025,
	title = {{Kimi-Audio Technical Report}},
	url = {http://arxiv.org/abs/2504.18425},
	doi = {10.48550/arXiv.2504.18425},
	number = {{arXiv}:2504.18425},
	publisher = {{arXiv}},
	author = {{KimiTeam} and Ding, Ding and Ju, Zeqian and Leng, Yichong and Liu, Songxiang and Liu, Tong and Shang, Zeyu and Shen, Kai and Song, Wei and Tan, Xu and Tang, Heyi and Wang, Zhengtao and Wei, Chu and Xin, Yifei and Xu, Xinran and Yu, Jianwei and Zhang, Yutao and Zhou, Xinyu and Charles, Y. and Chen, Jun and Chen, Yanru and Du, Yulun and He, Weiran and Hu, Zhenxing and Lai, Guokun and Li, Qingcheng and Liu, Yangyang and Sun, Weidong and Wang, Jianzhou and Wang, Yuzhi and Wu, Yuefeng and Wu, Yuxin and Yang, Dongchao and Yang, Hao and Yang, Ying and Yang, Zhilin and Yin, Aoxiong and Yuan, Ruibin and Zhang, Yutong and Zhou, Zaida},
	urldate = {2025-09-04},
	year = {2025},
	eprinttype = {arxiv},
	eprint = {2504.18425 [eess]}
}

@misc{chuQwen2AudioTechnicalReport2024,
	title = {{Qwen2-Audio Technical Report}},
	url = {http://arxiv.org/abs/2407.10759},
	doi = {10.48550/arXiv.2407.10759},
	number = {{arXiv}:2407.10759},
	publisher = {{arXiv}},
	author = {Chu, Yunfei and Xu, Jin and Yang, Qian and Wei, Haojie and Wei, Xipin and Guo, Zhifang and Leng, Yichong and Lv, Yuanjun and He, Jinzheng and Lin, Junyang and Zhou, Chang and Zhou, Jingren},
	urldate = {2025-09-04},
	year = {2024},
	eprinttype = {arxiv},
	eprint = {2407.10759 [eess]},
}

@misc{yourtts,
      title={{YourTTS: Towards Zero-Shot Multi-Speaker TTS and Zero-Shot Voice Conversion for everyone}}, 
      author={Edresson Casanova and Julian Weber and Christopher Shulby and Arnaldo Candido Junior and Eren Gölge and Moacir Antonelli Ponti},
      year={2023},
      eprint={2112.02418},
      archivePrefix={arXiv},
      primaryClass={cs.SD},
      url={https://arxiv.org/abs/2112.02418}, 
}

@misc{liuAudioMarkBenchBenchmarkingRobustness,
      title={{AudioMarkBench: Benchmarking Robustness of Audio Watermarking}}, 
      author={Hongbin Liu and Moyang Guo and Zhengyuan Jiang and Lun Wang and Neil Zhenqiang Gong},
      year={2024},
      eprint={2406.06979},
      archivePrefix={arXiv},
      primaryClass={cs.LG},
      url={https://arxiv.org/abs/2406.06979}, 
}

@inproceedings{AudioMarkNet,
author = {Zong, Wei and Chow, Yang-Wai and Susilo, Willy and Baek, Joonsang and Camtepe, Seyit},
title = {{AudioMarkNet: audio watermarking for deepfake speech detection}},
year = {2025},
isbn = {978-1-939133-52-6},
publisher = {USENIX Association},
address = {USA},
booktitle = {Proceedings of the 34th USENIX Conference on Security Symposium},
articleno = {240},
numpages = {20},
location = {Seattle, WA, USA},
series = {SEC '25}
}

@misc{romanLatentWatermarkingAudio2025,
      title={{Latent Watermarking of Audio Generative Models}}, 
      author={Robin San Roman and Pierre Fernandez and Antoine Deleforge and Yossi Adi and Romain Serizel},
      year={2024},
      eprint={2409.02915},
      archivePrefix={arXiv},
      primaryClass={cs.SD},
      url={https://arxiv.org/abs/2409.02915}, 
}

@misc{chenWavMarkWatermarkingAudio2024,
	title = {{WavMark}: Watermarking for Audio Generation},
	url = {http://arxiv.org/abs/2308.12770},
	doi = {10.48550/arXiv.2308.12770},
	shorttitle = {{WavMark}},
	number = {{arXiv}:2308.12770},
	publisher = {{arXiv}},
	author = {Chen, Guangyu and Wu, Yu and Liu, Shujie and Liu, Tao and Du, Xiaoyong and Wei, Furu},
	urldate = {2025-09-05},
    year = {2024},
	eprinttype = {arxiv},
	eprint = {2308.12770 [cs]},
}

@misc{liuGROOTGeneratingRobust2024,
      title={{GROOT: Generating Robust Watermark for Diffusion-Model-Based Audio Synthesis}}, 
      author={Weizhi Liu and Yue Li and Dongdong Lin and Hui Tian and Haizhou Li},
      year={2024},
      eprint={2407.10471},
      archivePrefix={arXiv},
      primaryClass={cs.CR},
      url={https://arxiv.org/abs/2407.10471}, 
}

@misc{ozerComprehensiveRealWorldAssessment2025,
	title = {{A Comprehensive Real-World Assessment of Audio Watermarking Algorithms: Will They Survive Neural Codecs?}},
	url = {http://arxiv.org/abs/2505.19663},
	doi = {10.48550/arXiv.2505.19663},
	shorttitle = {A Comprehensive Real-World Assessment of Audio Watermarking Algorithms},
	number = {{arXiv}:2505.19663},
	publisher = {{arXiv}},
	author = {Özer, Yigitcan and Choi, Woosung and Serrà, Joan and Singh, Mayank Kumar and Liao, Wei-Hsiang and Mitsufuji, Yuki},
	urldate = {2025-11-13},
	year = {2025},
	eprinttype = {arxiv},
	eprint = {2505.19663 [cs]},
}

@misc{liuDetectingVoiceCloning2023,
	title = {{Detecting Voice Cloning Attacks via Timbre Watermarking}},
	url = {http://arxiv.org/abs/2312.03410},
	doi = {10.48550/arXiv.2312.03410},
	number = {{arXiv}:2312.03410},
	publisher = {{arXiv}},
	author = {Liu, Chang and Zhang, Jie and Zhang, Tianwei and Yang, Xi and Zhang, Weiming and Yu, Nenghai},
	urldate = {2025-09-23},
	year = {2023},
	eprinttype = {arxiv},
	eprint = {2312.03410 [cs]},
}

@ARTICLE{spreadspectral,
  author={Kirovski, D. and Malvar, H.S.},
  journal={IEEE Transactions on Signal Processing}, 
  title={{Spread-spectrum watermarking of audio signals}}, 
  year={2003},
  volume={51},
  number={4},
  pages={1020-1033},
  doi={10.1109/TSP.2003.809384}
}

@misc{oreillyDeepAudioWatermarks2025,
	title = {{Deep Audio Watermarks are Shallow: Limitations of Post-Hoc Watermarking Techniques for Speech}},
	url = {http://arxiv.org/abs/2504.10782},
	doi = {10.48550/arXiv.2504.10782},
	shorttitle = {Deep Audio Watermarks are Shallow},
	number = {{arXiv}:2504.10782},
	publisher = {{arXiv}},
	author = {O'Reilly, Patrick and Jin, Zeyu and Su, Jiaqi and Pardo, Bryan},
	year = {2025},
	eprinttype = {arxiv},
	eprint = {2504.10782 [cs]},
}

@article{swansonRobustAudioWatermarking1998,
  title     = "Robust audio watermarking using perceptual masking",
  author    = "Swanson, Mitchell D and Zhu, Bin and Tewfik, Ahmed H and Boney,
               Laurence",
  journal   = "Signal Processing",
  publisher = "Elsevier BV",
  volume    =  66,
  number    =  3,
  pages     = "337--355",
  month     =  may,
  year      =  1998,
  language  = "en"
}

@INPROCEEDINGS{singhSilentCipherDeepAudio2024,
  title      = "{SilentCipher}: Deep Audio Watermarking",
  booktitle  = "Interspeech 2024",
  author     = "Singh, Mayank Kumar and Takahashi, Naoya and Liao, Weihsiang
                and Mitsufuji, Yuki",
  publisher  = "ISCA",
  pages      = "2235--2239",
  month      =  sep,
  year       =  2024,
  address    = "ISCA",
  conference = "Interspeech 2024"
}

@misc{librispeech,
  title           = "Librispeech: An {ASR} corpus based on public domain audio
                     books",
  booktitle       = "2015 {IEEE} International Conference on Acoustics, Speech
                     and Signal Processing ({ICASSP})",
  author          = "Panayotov, Vassil and Chen, Guoguo and Povey, Daniel and
                     Khudanpur, Sanjeev",
  publisher       = "IEEE",
  month           =  apr,
  year            =  2015,
  conference      = "ICASSP 2015 - 2015 IEEE International Conference on
                     Acoustics, Speech and Signal Processing (ICASSP)",
  location        = "South Brisbane, Queensland, Australia"
}

@misc{RIAA_2024_Suno_Udio,
  author       = {{Recording Industry Association of America (RIAA)}},
  title        = {{Record Companies Bring Landmark Cases for Responsible AI Against Suno and Udio in Boston and New York Federal Courts, Respectively}},
  year         = {2024},
  month        = jun,
  howpublished = {\href{https://www.riaa.com/record-companies-bring-landmark-cases-for-responsible-ai-against-suno-and-udio-in-boston-and-new-york-federal-courts-respectively/}{RIAA Press Release}},
}

@misc{whisper,
      title={{Robust Speech Recognition via Large-Scale Weak Supervision}}, 
      author={Alec Radford and Jong Wook Kim and Tao Xu and Greg Brockman and Christine McLeavey and Ilya Sutskever},
      year={2022},
      eprint={2212.04356},
      archivePrefix={arXiv},
      primaryClass={eess.AS},
      url={https://arxiv.org/abs/2212.04356}, 
}

@misc{wang2025sparkttsefficientllmbasedtexttospeech,
      title={{Spark-TTS: An Efficient LLM-Based Text-to-Speech Model with Single-Stream Decoupled Speech Tokens}}, 
      author={Xinsheng Wang and Mingqi Jiang and Ziyang Ma and Ziyu Zhang and Songxiang Liu and Linqin Li and Zheng Liang and Qixi Zheng and Rui Wang and Xiaoqin Feng and Weizhen Bian and Zhen Ye and Sitong Cheng and Ruibin Yuan and Zhixian Zhao and Xinfa Zhu and Jiahao Pan and Liumeng Xue and Pengcheng Zhu and Yunlin Chen and Zhifei Li and Xie Chen and Lei Xie and Yike Guo and Wei Xue},
      year={2025},
      eprint={2503.01710},
      archivePrefix={arXiv},
      primaryClass={cs.SD},
      url={https://arxiv.org/abs/2503.01710}, 
}

@misc{guardian2024wppdeepfake,
  title        = {{CEO of world’s biggest ad firm targeted by deepfake scam}},
  publisher      = {The Guardian},
  year         = {2024},
  url          = {https://www.theguardian.com/technology/article/2024/may/10/ceo-wpp-deepfake-scam}
}

@misc{brewster2021voiceclone35m,
  title        = {{Fraudsters Cloned Company Director’s Voice In \$35 Million Heist}},
  author      = {Forbes},
  year         = {2021},
  url          = {https://www.forbes.com/sites/thomasbrewster/2021/10/14/huge-bank-fraud-uses-deep-fake-voice-tech-to-steal-millions/}
}

@misc{guardian2025aigeneratedmusic,
  publisher    = {{The Guardian}},
  title     = {{An AI-generated band got 1m plays on Spotify. Now music insiders say listeners should be warned}},
  year      = {2025},
  howpublished = {\href{https://www.theguardian.com/technology/2025/jul/14/an-ai-generated-band-got-1m-plays-on-spotify-now-music-insiders-say-listeners-should-be-warned}{The Guardian (Tech)}},
}

@misc{andriushchenkoSquareAttackQueryefficient2020,
      title={{Square Attack: a query-efficient black-box adversarial attack via random search}}, 
      author={Maksym Andriushchenko and Francesco Croce and Nicolas Flammarion and Matthias Hein},
      year={2020},
      eprint={1912.00049},
      archivePrefix={arXiv},
      primaryClass={cs.LG},
      url={https://arxiv.org/abs/1912.00049}, 
}

@ARTICLE{park2024ai,
  title     = "{AI} deception: A survey of examples, risks, and potential
               solutions",
  author    = "Park, Peter S and Goldstein, Simon and O'Gara, Aidan and Chen,
               Michael and Hendrycks, Dan",
  journal   = "Patterns (N. Y.)",
  publisher = "Elsevier BV",
  volume    =  5,
  number    =  5,
  pages     = "100988",
  month     =  may,
  year      =  2024,
  copyright = "http://creativecommons.org/licenses/by-nc-nd/4.0/",
  language  = "en"
}

@article{STFT,
author = {D. Gabor },
title = {{Theory of communication. Part 1: The analysis of information}},
journal = {Journal of the Institution of Electrical Engineers - Part III: Radio and Communication Engineering},
volume = {93},
issue = {26},
pages = {429-441},
year = {1946},
doi = {10.1049/ji-3-2.1946.0074},

URL = {https://digital-library.theiet.org/doi/abs/10.1049/ji-3-2.1946.0074},
eprint = {https://digital-library.theiet.org/doi/pdf/10.1049/ji-3-2.1946.0074}
}

@article{guardian2024hongkongdeepfake,
  author  = {Milmo, Dan},
  title   = {{Company worker in Hong Kong pays out \pounds20m in deepfake video call scam}},
  journal = {The Guardian},
  year    = {2024},
  month   = {2},
  day     = {5},
  url     = {https://www.theguardian.com/world/2024/feb/05/hong-kong-company-deepfake-video-conference-call-scam}
}

@misc{hu2026qwen3ttstechnicalreport,
      title={{Qwen3-TTS Technical Report}}, 
      author={Hangrui Hu and Xinfa Zhu and Ting He and Dake Guo and Bin Zhang and Xiong Wang and Zhifang Guo and Ziyue Jiang and Hongkun Hao and Zishan Guo and Xinyu Zhang and Pei Zhang and Baosong Yang and Jin Xu and Jingren Zhou and Junyang Lin},
      year={2026},
      eprint={2601.15621},
      archivePrefix={arXiv},
      primaryClass={cs.SD},
      url={https://arxiv.org/abs/2601.15621}, 
}

@misc{yamagishi2019vctk,
  author={Yamagishi, Junichi and Veaux, Christophe and MacDonald, Kirsten},
  title={{CSTR VCTK Corpus}: English Multi-speaker Corpus for {CSTR} Voice Cloning Toolkit (version 0.92)},
  publisher={University of Edinburgh. The Centre for Speech Technology Research (CSTR)},
  year=2019,
  doi={10.7488/ds/2645},
}

@ARTICLE{clap,
  title         = "{CLAP}: Learning audio concepts from natural language
                   supervision",
  author        = "Elizalde, Benjamin and Deshmukh, Soham and Al Ismail,
                   Mahmoud and Wang, Huaming",
  month         =  jun,
  year          =  2022,
  copyright     = "http://creativecommons.org/licenses/by/4.0/",
  archivePrefix = "arXiv",
  primaryClass  = "cs.SD",
  eprint        = "2206.04769"
}

@ARTICLE{musicGen,
  title         = {{Simple and Controllable Music Generation}},
  author        = "Copet, Jade and Kreuk, Felix and Gat, Itai and Remez, Tal
                   and Kant, David and Synnaeve, Gabriel and Adi, Yossi and
                   D{\'e}fossez, Alexandre",
  month         =  jun,
  year          =  2023,
  copyright     = "http://arxiv.org/licenses/nonexclusive-distrib/1.0/",
  archivePrefix = "arXiv",
  primaryClass  = "cs.SD",
  eprint        = "2306.05284"
}

@ARTICLE{audioldm2,
  title         = {{AudioLDM 2: Learning Holistic Audio Generation with Self-supervised Pretraining}},
  author        = "Liu, Haohe and Yuan, Yi and Liu, Xubo and Mei, Xinhao and
                   Kong, Qiuqiang and Tian, Qiao and Wang, Yuping and Wang,
                   Wenwu and Wang, Yuxuan and Plumbley, Mark D",
  month         =  aug,
  year          =  2023,
  copyright     = "http://arxiv.org/licenses/nonexclusive-distrib/1.0/",
  archivePrefix = "arXiv",
  primaryClass  = "cs.SD",
  eprint        = "2308.05734"
}

\appendices
\section{Hyperparameter Choices for the Distortions} \label{app:hyper}

We evaluate robustness under a comprehensive suite of audio distortions. For Gaussian noise, we use additive white Gaussian noise at SNRs of 20 dB and 30 dB. For resampling, we downsample the audio to 50\% of the original sampling rate and then upsample it back to the original rate. 
For volume scaling, we multiply the waveform amplitude by fixed factors of 0.5 and 2.0. For low-pass and high-pass filtering, we use an FFT-domain low-pass filter with a 4 kHz cutoff and an FFT-domain high-pass filter with a 300 Hz cutoff. 
For echo, we add a single delayed copy with a delay of 0.1s with a decay factor of 0.4. 
For clipping, we hard-clip the waveform to the range $[-0.5, 0.5]$.
For quantization, we quantize the waveform to 8-bit precision. For phase shift, we perturb the phase of each FFT bin with a random shift sampled uniformly from $[-0.5, 0.5]$ radians.

\begin{table}[ht!]
\centering
\setlength{\tabcolsep}{4pt}

\caption{Comparison of average character and word similarity scores using different Whisper captioning models~\cite{whisper} on AudioSeal samples and \tool samples.}
\begin{tabular}{lcccc}
\toprule
\multirow{2}{*}{\textbf{Whisper}} 
& \multicolumn{2}{c}{\textbf{Char Similarity}} 
& \multicolumn{2}{c}{\textbf{Word Similarity}} \\
\cmidrule(lr){2-3} \cmidrule(lr){4-5}
& \textbf{AudioSeal} & \textbf{\tool}
& \textbf{AudioSeal} & \textbf{\tool} \\
\midrule
base   & 0.9952 & 0.9808 & 0.9870 & 0.9463 \\
tiny   & 0.9967 & 0.9817 & 0.9845 & 0.9274 \\
small  & 0.9914 & 0.9888 & 0.9781 & 0.9569 \\
medium & 0.9782 & 0.9876 & 0.9700 & 0.9661 \\
large  & 0.9919 & 0.9865 & 0.9769 & 0.9644 \\
\midrule
\textbf{Average} 
& \textbf{0.9907} & \textbf{0.9851}
& \textbf{0.9793} & \textbf{0.9522} \\
\bottomrule
\end{tabular}
\label{tab:captioning_similarity}
\end{table}
\section{Watermarked Audio Intelligibility Interpreted by Whisper Captioning Models}\label{app:caption}
Table~\ref{tab:captioning_similarity} shows that watermarked audio produced by \tool shows high character similarity and word similarity, comparable to the most competitive AudioSeal baseline regardless of captioning model size, indicating high intelligibility and preservation of linguistic content despite the latent-space semantic shifts induced by watermark embedding.

\begin{table*}[th!]
\centering
\scriptsize
\renewcommand{\arraystretch}{1.15}
\caption{Robustness and fidelity evaluation of \tool under different audio distortions on the LibriSpeech dataset.}
\label{tab:distortion_results_detail_lambda_libri}
\begin{tabular}{lcccccc|cc}
\toprule
\textbf{Distortion}
& \textbf{\detacc~(\%)} & \textbf{\bitacc~(\%)}
& \textbf{P~(\%)} & \textbf{R~(\%)} & \textbf{F1~(\%)}
& \textbf{AUC} & \textbf{ViSQOL} & \textbf{NISQA} \\
\midrule
\textbf{None} & 100.00 & 99.75 & 100.00 & 100.00 & 100.00 & 1.000 & 4.44 & 4.66 \\
Gaussian 20dB & 100.00 & 97.53 & 100.00 & 100.00 & 100.00 & 1.000 & 3.58 & 2.47 \\
Gaussian 30dB & 99.50 & 98.93 & 100.00 & 99.00 & 99.50 & 0.995 & 4.03 & 3.27 \\
Resampling & 100.00 & 99.31 & 100.00 & 100.00 & 100.00 & 1.000 & 4.29 & 4.46 \\
Volume 0.5x & 100.00 & 99.68 & 100.00 & 100.00 & 100.00 & 1.000 & 4.43 & 4.60 \\
Volume 2x & 100.00 & 99.71 & 100.00 & 100.00 & 100.00 & 1.000 & 4.43 & 4.41 \\
Low-pass & 100.00 & 90.97 & 100.00 & 100.00 & 100.00 & 1.000 & 3.86 & 3.69 \\
High-pass & 100.00 & 99.44 & 100.00 & 100.00 & 100.00 & 1.000 & 3.31 & 2.52 \\
Echo & 100.00 & 96.81 & 100.00 & 100.00 & 100.00 & 1.000 & 3.61 & 2.90 \\
Clipping & 100.00 & 99.75 & 100.00 & 100.00 & 100.00 & 1.000 & 4.43 & 4.61 \\
Quantization & 100.00 & 99.06 & 100.00 & 100.00 & 100.00 & 1.000 & 4.12 & 2.72 \\
Phase Shift & 99.50 & 94.34 & 99.01 & 100.00 & 99.50 & 0.995 & 3.23 & 2.23 \\
\midrule
\textbf{Average} & 99.92 & 97.94 & 99.92 & 99.92 & 99.92 & 0.999 & 3.98 & 3.55 \\

\bottomrule
\end{tabular}
\end{table*}

\begin{table*}[th!]
\centering
\scriptsize
\caption{Robustness and fidelity evaluation of \tool under different audio distortions on the VCTK dataset. \tool is trained on the LibriSpeech dataset and transferred to the VCTK dataset in this setting.}
\begin{tabular}{lcccccc|cc}
\toprule
\textbf{Distortion} & \textbf{\detacc~(\%)} & \textbf{\bitacc~(\%)} & \textbf{P~(\%)} &\textbf{ R~(\%) }& \textbf{F1~(\%)} & \textbf{AUC} & \textbf{ViSQOL} & \textbf{NISQA} \\
\midrule
\textbf{None}   & 100.00 & 98.13 & 100.00 & 100.00 & 100.00 & 1.000 & 4.26 & 4.81 \\
Gaussian 20dB  & 99.50  & 92.81 & 99.01  & 100.00 & 99.50  & 0.995 & 3.33 & 2.93 \\
Gaussian 30dB  & 100.00 & 96.44 & 100.00 & 100.00 & 100.00 & 1.000 & 3.68 & 3.32 \\
Resampling     & 100.00 & 96.72 & 100.00 & 100.00 & 100.00 & 1.000 & 4.12 & 4.65 \\
Volume 0.5x    & 100.00 & 98.09 & 100.00 & 100.00 & 100.00 & 1.000 & 4.26 & 4.77 \\
Volume 2x      & 100.00 & 98.25 & 100.00 & 100.00 & 100.00 & 1.000 & 4.26 & 4.54 \\
Low-pass       & 100.00 & 89.16 & 100.00 & 100.00 & 100.00 & 1.000 & 3.69 & 3.89 \\
High-pass      & 100.00 & 96.72 & 100.00 & 100.00 & 100.00 & 1.000 & 3.15 & 2.56 \\
Echo           & 100.00 & 93.53 & 100.00 & 100.00 & 100.00 & 1.000 & 3.37 & 3.13 \\
Clipping       & 100.00 & 98.06 & 100.00 & 100.00 & 100.00 & 1.000 & 4.25 & 4.78 \\
Quantization   & 100.00 & 95.66 & 100.00 & 100.00 & 100.00 & 1.000 & 3.63 & 1.99 \\
Phase Shift    & 99.00  & 87.34 & 98.04  & 100.00 & 99.01  & 0.990 & 2.86 & 2.41 \\
\midrule
\textbf{Average} & 99.88 & 95.08 & 99.75 & 100.00 & 99.88 & 0.999 & 3.74 & 3.65 \\
\bottomrule
\end{tabular}
\label{tab:tool_vctk}
\end{table*}

\begin{table*}[th!]
\centering
\scriptsize
\caption{Robustness and fidelity evaluation of AudioSeal under different audio distortions on the LibriSpeech dataset.}
\begin{tabular}{lcccccc|cc}
\toprule
\textbf{Distortion} & \textbf{\detacc~(\%)} & \textbf{\bitacc~(\%)} & \textbf{P~(\%)} &\textbf{ R~(\%) }& \textbf{F1~(\%)} & \textbf{AUC} & \textbf{ViSQOL} & \textbf{NISQA} \\
\midrule
\textbf{None}   & 100.00 & 51.94 & 100.00 & 100.00 & 100.00 & 1.000 & 4.94 & 4.49 \\
Gaussian 20dB  & 52.00  & 50.75 & 100.00 & 4.00   & 7.69   & 0.520 & 3.59 & 1.87 \\
Gaussian 30dB  & 94.00  & 50.00 & 100.00 & 88.00  & 93.62  & 0.940 & 4.25 & 2.98 \\
Resampling     & 100.00 & 49.94 & 100.00 & 100.00 & 100.00 & 1.000 & 4.19 & 3.92 \\
Volume 0.5x    & 100.00 & 49.94 & 100.00 & 100.00 & 100.00 & 1.000 & 4.94 & 4.61 \\
Volume 2x      & 100.00 & 49.94 & 100.00 & 100.00 & 100.00 & 1.000 & 4.94 & 4.07 \\
Low-pass       & 100.00 & 49.94 & 100.00 & 100.00 & 100.00 & 1.000 & 4.22 & 3.73 \\
High-pass      & 100.00 & 49.94 & 100.00 & 100.00 & 100.00 & 1.000 & 3.66 & 2.70 \\
Echo           & 100.00 & 49.94 & 100.00 & 100.00 & 100.00 & 1.000 & 3.89 & 2.89 \\
Clipping       & 100.00 & 49.94 & 100.00 & 100.00 & 100.00 & 1.000 & 4.82 & 4.33 \\
Quantization          & 95.50  & 49.88 & 100.00 & 91.00  & 95.29  & 0.955 & 4.50 & 3.31 \\
Phase Shift    & 52.00  & 50.50 & 100.00 & 4.00   & 7.69   & 0.520 & 3.41 & 2.41 \\
\midrule
\textbf{Average} & 91.13 & 50.22 & 100.00 & 82.25 & 83.69 & 0.911 & 4.28 & 3.44 \\
\bottomrule
\end{tabular}
\label{tab:audioseal_distortion}
\end{table*}

\begin{table*}[th!]
\centering
\scriptsize
\caption{Robustness and fidelity evaluation of WavMark under different audio distortions on the LibriSpeech dataset.}
\begin{tabular}{lcccccc|cc}
\toprule
\textbf{Distortion} & \textbf{\detacc~(\%)} & \textbf{\bitacc~(\%)} & \textbf{P~(\%)} &\textbf{ R~(\%) }& \textbf{F1~(\%)} & \textbf{AUC} & \textbf{ViSQOL} & \textbf{NISQA} \\
\midrule
\textbf{None}   & 100.00 & 100.00 & 100.00 & 100.00 & 100.00 & 1.000 & 4.66 & 4.52 \\
Gaussian 20dB  & 50.50  & 0.88   & 100.00 & 1.00   & 1.98   & 0.505 & 3.61 & 1.92 \\
Gaussian 30dB  & 85.00  & 67.25  & 100.00 & 70.00  & 82.35  & 0.850 & 4.25 & 3.04 \\
Resampling     & 100.00 & 100.00 & 100.00 & 100.00 & 100.00 & 1.000 & 4.04 & 4.00 \\
Volume 0.5x    & 100.00 & 100.00 & 100.00 & 100.00 & 100.00 & 1.000 & 4.66 & 4.67 \\
Volume 2x      & 100.00 & 100.00 & 100.00 & 100.00 & 100.00 & 1.000 & 4.66 & 4.13 \\
Low-pass       & 100.00 & 100.00 & 100.00 & 100.00 & 100.00 & 1.000 & 4.06 & 3.84 \\
High-pass      & 100.00 & 100.00 & 100.00 & 100.00 & 100.00 & 1.000 & 3.48 & 2.76 \\
Echo           & 100.00 & 99.88  & 100.00 & 100.00 & 100.00 & 1.000 & 3.74 & 2.97 \\
Clipping       & 100.00 & 100.00 & 100.00 & 100.00 & 100.00 & 1.000 & 4.55 & 4.39 \\
Quantization          & 96.50  & 91.94  & 100.00 & 93.00  & 96.37  & 0.965 & 4.44 & 3.33 \\
Phase Shift    & 59.00  & 15.94  & 100.00 & 18.00  & 30.51  & 0.590 & 3.42 & 2.41 \\
\midrule
\textbf{Average} & 90.92 & 81.32 & 100.00 & 81.83 & 84.27 & 0.909 & 4.13 & 3.50 \\
\bottomrule
\end{tabular}
\label{tab:wavmark_distortion}
\end{table*}

\begin{table*}[th!]
\centering
\scriptsize
\caption{Robustness and fidelity evaluation of AudioMarkNet under different audio distortions on the VCTK dataset. 
AudioMarkNet is trained and evaluated only on the VCTK dataset.}
\begin{tabular}{lcccccc|cc}
\toprule
\textbf{Distortion} & \textbf{\detacc~(\%)} & \textbf{\bitacc~(\%)} & \textbf{P~(\%)} &\textbf{ R~(\%) }& \textbf{F1~(\%)} & \textbf{AUC} & \textbf{ViSQOL} & \textbf{NISQA} \\
\midrule
\textbf{None}   & 93.50 & 83.54 & 100.00 & 87.00 & 93.05 & 0.935 & 4.60 & 4.54 \\
Gaussian 20dB  & 83.50 & 80.76 & 100.00 & 67.00 & 80.24 & 0.835 & 3.33 & 1.96 \\
Gaussian 30dB  & 93.50 & 84.15 & 100.00 & 87.00 & 93.05 & 0.935 & 3.81 & 3.07 \\
Resampling     & 93.50 & 83.54 & 100.00 & 87.00 & 93.05 & 0.935 & 3.92 & 3.71 \\
Volume 0.5x    & 93.50 & 83.34 & 100.00 & 87.00 & 93.05 & 0.935 & 4.60 & 4.50 \\
Volume 2x      & 93.00 & 83.24 & 100.00 & 86.00 & 92.47 & 0.930 & 4.60 & 4.29 \\
Low-pass       & 93.50 & 83.54 & 100.00 & 87.00 & 93.05 & 0.935 & 3.92 & 3.48 \\
High-pass      & 75.50 & 76.17 & 100.00 & 51.00 & 67.55 & 0.755 & 3.35 & 2.78 \\
Echo           & 88.50 & 82.56 & 100.00 & 77.00 & 87.01 & 0.885 & 3.58 & 2.91 \\
Clipping       & 93.50 & 83.54 & 100.00 & 87.00 & 93.05 & 0.935 & 4.59 & 4.53 \\
Quantization          & 73.50 & 74.38 & 100.00 & 47.00 & 63.95 & 0.735 & 3.70 & 1.68 \\
Phase Shift    & 51.50 & 48.33 & 71.43  & 5.00  & 9.35  & 0.515 & 2.99 & 2.45 \\
\midrule
\textbf{Average} & 85.54 & 78.92 & 97.62 & 71.25 & 79.91 & 0.855 & 3.92 & 3.33 \\
\bottomrule
\end{tabular}
\label{tab:audiomarknet_distortion}
\end{table*}

\begin{table*}[th!]
\centering
\scriptsize
\caption{Robustness and fidelity evaluation of \tool with 16-bit watermark embedded on the LibriSpeech dataset.}
\begin{tabular}{lcccccc|cc}
\toprule
\textbf{Distortion} & \textbf{\detacc~(\%)} & \textbf{\bitacc~(\%)} & \textbf{P~(\%)} &\textbf{ R~(\%) }& \textbf{F1~(\%)} & \textbf{AUC} & \textbf{ViSQOL} & \textbf{NISQA} \\
\midrule
\textbf{None}   & 100.00 & 99.88 & 100.00 & 100.00 & 100.00 & 1.000 & 4.52 & 4.68 \\
Gaussian 20dB  & 99.00  & 98.12 & 100.00 & 98.00  & 98.99  & 0.990 & 3.60 & 2.41 \\
Gaussian 30dB  & 99.50  & 99.00 & 100.00 & 99.00  & 99.50  & 0.995 & 4.08 & 3.27 \\
Resampling     & 100.00 & 99.88 & 100.00 & 100.00 & 100.00 & 1.000 & 4.36 & 4.51 \\
Volume 0.5x    & 100.00 & 99.81 & 100.00 & 100.00 & 100.00 & 1.000 & 4.52 & 4.69 \\
Volume 2x      & 100.00 & 99.88 & 100.00 & 100.00 & 100.00 & 1.000 & 4.52 & 4.43 \\
Low-pass       & 100.00 & 98.94 & 100.00 & 100.00 & 100.00 & 1.000 & 3.94 & 3.75 \\
High-pass      & 100.00 & 99.62 & 100.00 & 100.00 & 100.00 & 1.000 & 3.40 & 2.66 \\
Echo           & 100.00 & 99.25 & 100.00 & 100.00 & 100.00 & 1.000 & 3.67 & 2.85 \\
Clipping       & 100.00 & 99.88 & 100.00 & 100.00 & 100.00 & 1.000 & 4.50 & 4.62 \\
Quantization          & 100.00 & 99.38 & 100.00 & 100.00 & 100.00 & 1.000 & 4.20 & 2.86 \\
Phase Shift    & 100.00 & 97.62 & 100.00 & 100.00 & 100.00 & 1.000 & 3.27 & 2.22 \\
\midrule
\textbf{Average} & 99.88 & 99.27 & 100.00 & 99.75 & 99.87 & 0.999 & 4.05 & 3.58 \\
\bottomrule
\end{tabular}
\label{tab:16bits}
\end{table*}

\begin{table*}[ht!]
\centering
\scriptsize
\caption{Robustness and fidelity evaluation of \tool with 48-bit watermark embedded on the LibriSpeech dataset.}
\begin{tabular}{lcccccc|cc}
\toprule
\textbf{Distortion} & \textbf{\detacc~(\%)} & \textbf{\bitacc~(\%)} & \textbf{P~(\%)} &\textbf{ R~(\%) }& \textbf{F1~(\%)} & \textbf{AUC} & \textbf{ViSQOL} & \textbf{NISQA} \\
\midrule
\textbf{None}   & 100.00 & 98.67 & 100.00 & 100.00 & 100.00 & 1.000 & 4.28 & 4.06 \\
Gaussian 20dB  & 100.00 & 96.17 & 100.00 & 100.00 & 100.00 & 1.000 & 3.55 & 2.49 \\
Gaussian 30dB  & 99.00 & 98.23 & 100.00 & 98.00 & 98.99 & 0.990 & 3.95 & 3.24 \\
Resampling     & 100.00 & 95.98 & 100.00 & 100.00 & 100.00 & 1.000 & 4.12 & 3.76 \\
Volume 0.5x    & 100.00 & 98.63 & 100.00 & 100.00 & 100.00 & 1.000 & 4.28 & 4.01 \\
Volume 2x      & 100.00 & 98.71 & 100.00 & 100.00 & 100.00 & 1.000 & 4.28 & 3.82 \\
Low-pass       & 100.00 & 91.88 & 100.00 & 100.00 & 100.00 & 1.000 & 3.77 & 3.10 \\
High-pass      & 100.00 & 97.98 & 100.00 & 100.00 & 100.00 & 1.000 & 3.16 & 2.36 \\
Echo           & 100.00 & 96.46 & 100.00 & 100.00 & 100.00 & 1.000 & 3.50 & 2.27 \\
Clipping       & 100.00 & 98.67 & 100.00 & 100.00 & 100.00 & 1.000 & 4.28 & 4.04 \\
Quantization          & 100.00 & 97.19 & 100.00 & 100.00 & 100.00 & 1.000 & 4.00 & 2.49 \\
Phase Shift    & 99.50 & 95.67 & 100.00 & 99.00 & 99.50 & 0.995 & 3.16 & 2.17 \\
\midrule
\textbf{Average} & 99.88 & 97.02 & 100.00 & 99.75 & 99.87 & 0.999 & 3.86 & 3.15 \\
\bottomrule
\end{tabular}
\label{tab:48bits}
\end{table*}

\begin{table*}[ht!]
\caption{\tool's robustness on \dadv generated by \madv after finetuning on watermarked LibriSpeech data (\dwm)}
\centering
\renewcommand{\arraystretch}{1.2}
\begin{tabular}{lllcccccc}
\toprule
\textbf{Attack} & \textbf{\madv} & \textbf{FT} & \textbf{\detacc~(\%)} & \textbf{\bitacc~(\%)} & \textbf{P~(\%)} & \textbf{R~(\%)} & \textbf{F1~(\%)} & \textbf{AUC} \\
\midrule

\multirow{5}{*}{Signal-level Distortions}
& YourTTS & Full & 99.17 & 87.77 & 100.00 & 98.33 & 99.15 & 0.992 \\\cmidrule{2-9}
& \multirow{2}{*}{SemanticVocoder} & Full & 99.96 & 93.62 & 100.00 & 99.92 & 99.96 & 1.000 \\
& & LoRA & 100.00 & 93.22 & 100.00 & 100.00 & 100.00 & 1.000 \\\cmidrule{2-9}
& \multirow{2}{*}{AudioLDM2} & Full & 94.96 & 81.90 & 100.00 & 89.92 & 94.58 & 0.999 \\
& & LoRA & 85.33 & 70.57 & 100.00 & 70.67 & 81.64 & 0.994 \\
\midrule

\multirow{5}{*}{\audiomarkbench}
& YourTTS & Full & 92.50 & 77.78 & 100.00 & 85.00 & 91.89 & 0.925 \\\cmidrule{2-9}
& \multirow{2}{*}{SemanticVocoder} & Full & 97.00 & 87.91 & 100.00 & 94.00 & 96.91 & 0.970 \\
& & LoRA & 99.50 & 88.09 & 100.00 & 99.00 & 99.50 & 0.995 \\\cmidrule{2-9}
& \multirow{2}{*}{AudioLDM2} & Full & 93.50 & 83.75 & 100.00 & 87.00 & 93.05 & 0.935 \\
& & LoRA & 71.50 & 67.25 & 100.00 & 43.00 & 60.14 & 0.715 \\
\midrule

\multirow{5}{*}{HarmonicAttack}
& YourTTS & Full & 100.00 & 78.63 & 100.00 & 100.00 & 100.00 & 1.000 \\\cmidrule{2-9}
& \multirow{2}{*}{SemanticVocoder} & Full & 99.00 & 85.25 & 99.00 & 99.00 & 99.00 & 1.000 \\
& & LoRA & 100.00 & 83.53 & 100.00 & 100.00 & 100.00 & 1.000 \\\cmidrule{2-9}
& \multirow{2}{*}{AudioLDM2} & Full &  94.00 & 75.47 & 97.83 & 90.00 & 93.75 & 0.990 \\
& & LoRA & 85.50 & 70.88 & 97.33 & 73.00 & 83.43 & 0.978 \\

\bottomrule
\end{tabular}
\label{tab:robustness_of_radioactivity_libri}
\end{table*}

\section{Detailed Results of \tool and Baseline Watermarks in No-Distortion Setting and against All 11 Distortion Settings} \label{app:distortion}
Tables~\ref{tab:distortion_results_detail_lambda_libri}, \ref{tab:audioseal_distortion}, and \ref{tab:wavmark_distortion} are the detailed results of \tool, AudioSeal, and WavMark on the LibriSpeech dataset under no-distortion and 11 different distortion settings, respectively.
Table~\ref{tab:audiomarknet_distortion} shows AudioMarkNet's performance on the VCTK dataset under common distortions. Since AudioMarkNet only supports VCTK, and to compare \tool with it, we record \tool's transfer results on VCTK under the same distortions in Table~\ref{tab:tool_vctk}, complementary to LibriSpeech results in Table~\ref{tab:distortion_results_detail_lambda_libri}.
All the averaged results are reported in Table~\ref{tab:distortion_results_comparison}.

\begin{figure*}[ht!]
    \centering
    \begin{subfigure}[t]{0.32\textwidth}
        \centering
        \includegraphics[width=\linewidth]{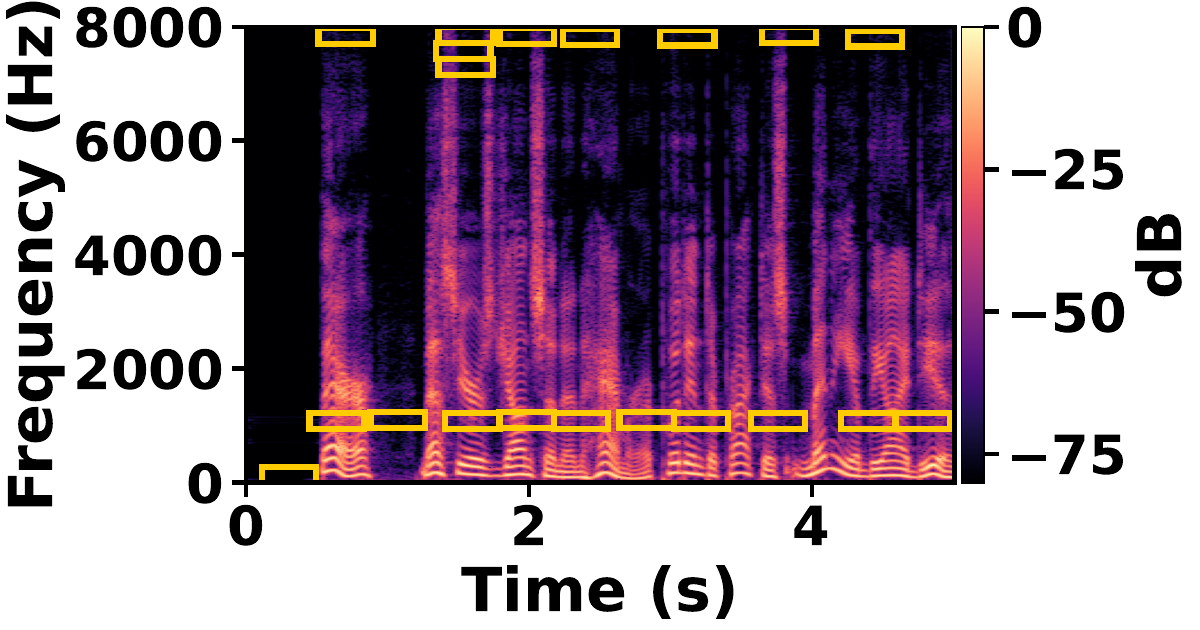}
        \caption{AudioSeal Sample 1.}
    \end{subfigure}
    \hfill
    \begin{subfigure}[t]{0.32\textwidth}
        \centering
        \includegraphics[width=\linewidth]{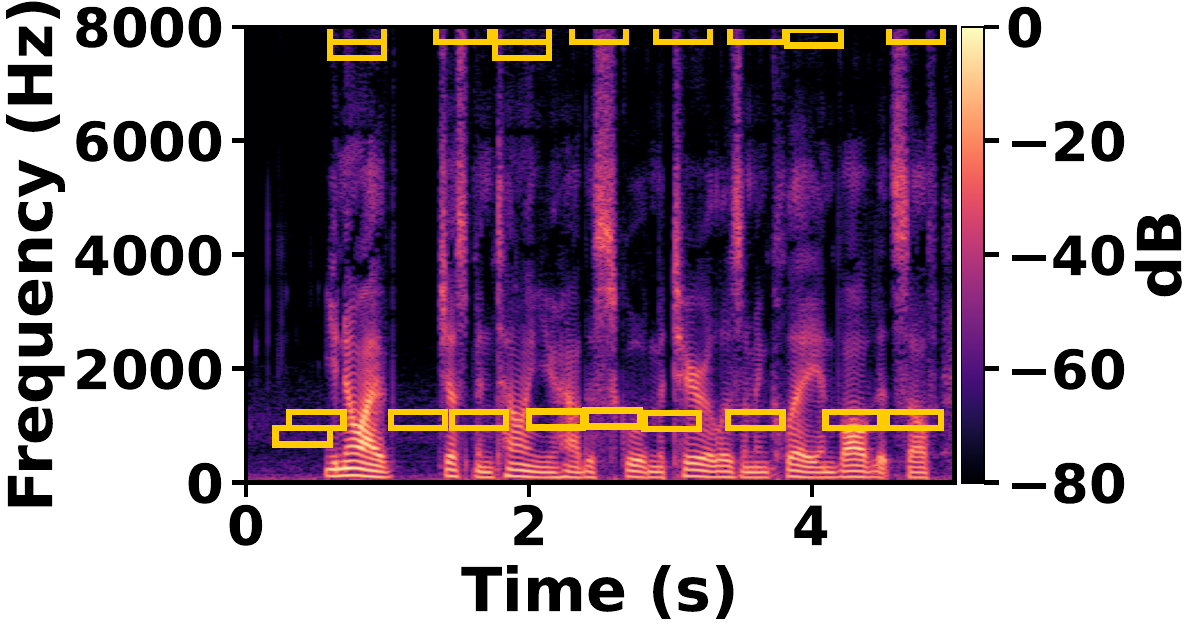}
        \caption{AudioSeal Sample 2.}
    \end{subfigure}
    \hfill
    \begin{subfigure}[t]{0.32\textwidth}
        \centering
        \includegraphics[width=\linewidth]{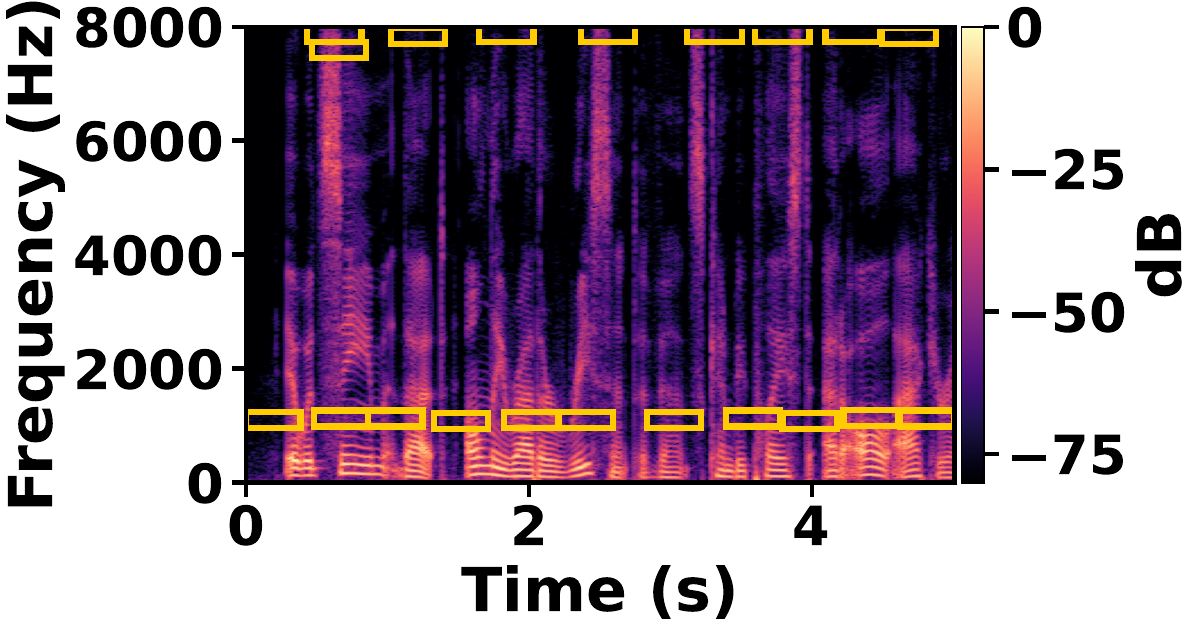}
        \caption{AudioSeal Sample 3.}
    \end{subfigure}

    \vspace{0.5em}

    \begin{subfigure}[t]{0.32\textwidth}
        \centering
        \includegraphics[width=\linewidth]{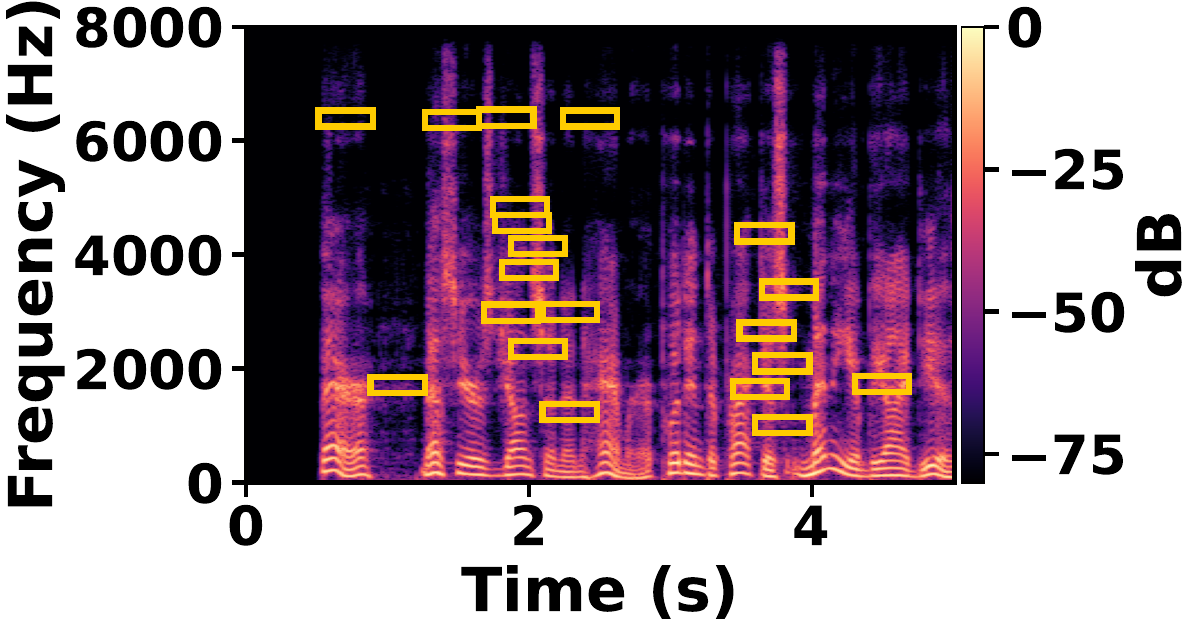}
        \caption{\tool Sample 1.}
    \end{subfigure}
    \hfill
    \begin{subfigure}[t]{0.32\textwidth}
        \centering
        \includegraphics[width=\linewidth]{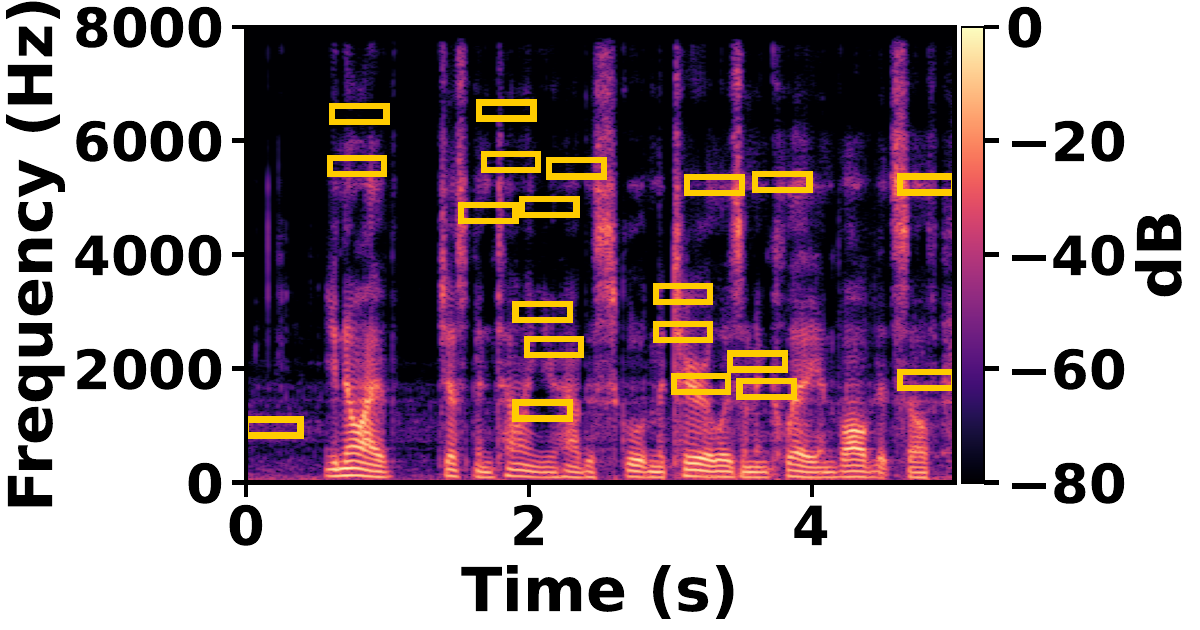}
        \caption{\tool Sample 2.}
    \end{subfigure}
    \hfill
    \begin{subfigure}[t]{0.32\textwidth}
        \centering
        \includegraphics[width=\linewidth]{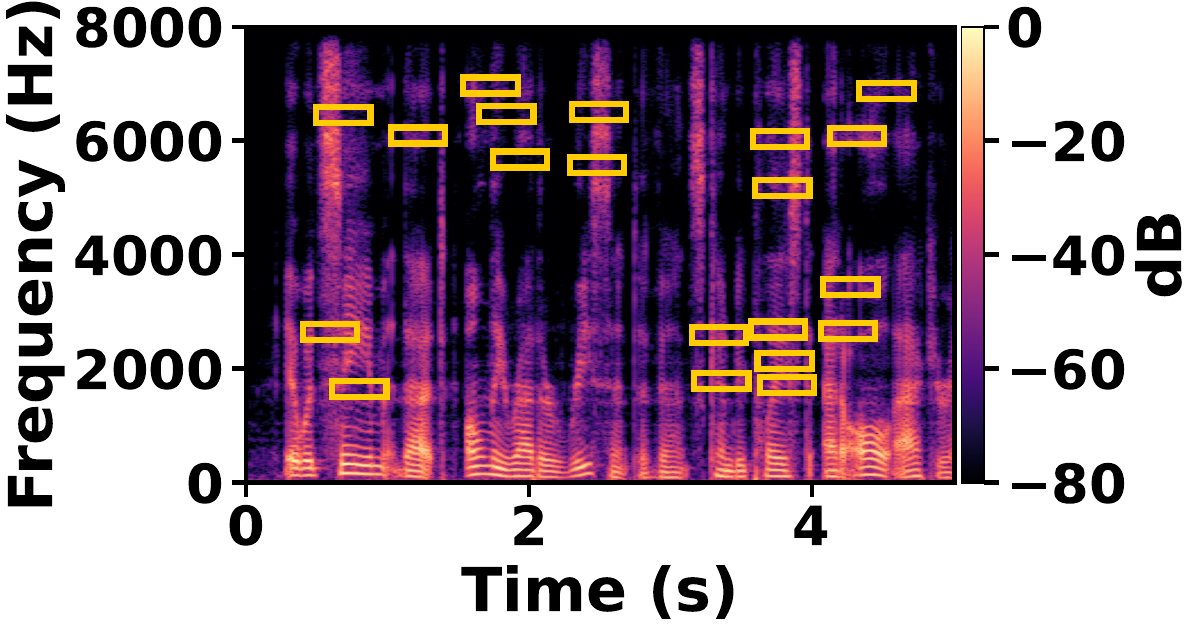}
        \caption{\tool Sample 3.}
    \end{subfigure}

    \caption{Linear-frequency STFT spectrogram visualizations of watermarked audio produced by \tool and AudioSeal across three representative LibriSpeech samples. Yellow boxes indicate the top-$K$ ($K=20$) residual regions, corresponding to the strongest residual regions in STFT space.}
    \label{fig:watermarked_spectrograms_stft}
\end{figure*}

\section{Detailed Results of \tool with Different Bit Sizes} \label{app:bit}
Complementary to the robustness results against distortions for the 32-bit watermark in Table~\ref{tab:distortion_results_detail_lambda_libri}, Table~\ref{tab:16bits} and Table~\ref{tab:48bits} show robustness results for 16-bit and 48-bit watermarks on the LibriSpeech dataset, respectively. The results suggest that LambdaMark remains robust across different bit sizes, with only minor performance variations.

\section{Watermark Robustness on Radioactive Samples} \label{app:robustness_of_radioactivity}
In addition to Table~\ref{tab:robustness_of_radioactivity_vctk_yourtts_compare} and Table~\ref{tab:robustness_of_radioactivity_vctk} demonstrating \tool's robust radioactivity on VCTK, compared to AudioMarkNet, which is only radioactive to YourTTS, we show \tool's robustness results on radioactive LibriSpeech samples (\dadv) in Table~\ref{tab:robustness_of_radioactivity_libri}. We observe similarly high robustness on LibriSpeech radioactive samples. It is worth noting that AudioMarkNet overfits to the VCTK dataset and the YourTTS audio generation model; it does not work on LibriSpeech (as indicated in their limitations) and is not compatible with SemanticVocoder and AudioLDM2 (See Table~\ref{table-downstream-models}). Therefore, we exclude its evaluation of robust radioactivity in Table~\ref{tab:robustness_of_radioactivity_vctk}~and Table~\ref{tab:robustness_of_radioactivity_libri}.

\section{STFT Spectrogram Diagrams}\label{app:spec}
Figure~\ref{fig:watermarked_spectrograms_stft} shows the STFT spectrograms complementary to mel spectrograms in Section~\ref{subsec:spectrogram}. The observations are consistent with Figure~\ref{fig:watermarked_spectrograms_mel}. It is worth noting that mel spectrograms compress and reweight frequencies using overlapping mel filters; a narrow high-frequency artifact may be strong in the STFT, but after mel filtering, it may be averaged out and no longer appear among the top-$K$ mel regions.

\end{document}